\documentclass[MSNbibl,nameyear,seceqn,dvips]{arxstspdf}
\usepackage{graphicx}
\usepackage{flushend}
\usepackage{stfloats}

\volume{27}
\issue{2}
\pubyear{2012}
\firstpage{202}
\lastpage{231}
\doi{10.1214/11-STS368}

\makeatletter

  \let\sv@tabnotetext\tabnotetext
  \let\sv@tabnotemark@fmt\tabnotemark@fmt
   \long\def\legend#1{{\let\tabnote@indent\leavevmode\sv@tabnotetext[]{}{#1}}}

\makeatother

\begin{document}
\begin{frontmatter}

\title{Statistical Significance of the~Netflix~Challenge}

\runtitle{Netflix Challenge}

\begin{aug}
\author{\fnms{Andrey} \snm{Feuerverger}\corref{}\ead[label=e1]{andrey@utstat.toronto.edu}},
\author{\fnms{Yu} \snm{He}\ead[label=e2]{njheyu@gmail.com}}
\and
\author{\fnms{Shashi} \snm{Khatri}\ead[label=e3]{contact@SpeakEmpirics.com}}

\runauthor{A. Feuerverger, Y. He and S. Khatri}

\affiliation{University of Toronto, Nanjing University and Speak Empirics Inc.}

\address{Andrey Feuerverger is Professor of Statistics,
Department of Statistics,
University of Toronto,
Toronto, Ontario, Canada M5S 3G3 \printead{e1}. Yu He is an Undergraduate Student,
Department of Mathematics,
Nanjing University
22 Hankou Road
Nanjing, 210093, China \printead{e2}.
Shashi Khatri is Director,
Speak Empirics,
1 Macklem Avenue,
Toronto, Ontario, Canada M6J 3M1  \printead{e3}.}

\end{aug}

\begin{abstract}
Inspired by the legacy of the Netflix contest, we provide an overview
of what has been learned---from our own efforts, and those of
others---concerning the problems of collaborative filtering and
recommender systems. The data set consists of about 100 million movie
ratings (from 1 to 5 stars) involving some 480 thousand users and some
18 thousand movies; the associated ratings matrix is about 99\% sparse.
The goal is to predict ratings that users will give to movies; systems
which can do this accurately have significant commercial applications,
particularly on the world wide web. We discuss, in some detail,
approaches to ``baseline'' modeling, singular value decomposition
(SVD), as well as kNN (nearest neighbor) and neural network models;
temporal effects, cross-validation issues, ensemble methods and other
considerations are discussed as well. We compare existing models in a
search for new models, and also discuss the mission-critical issues of
penalization and parameter shrinkage which arise when the dimensions of
a~parameter space reaches into the millions. Although much work on such
problems has been carried out by the computer science and machine
learning communities, our goal here is to address a statistical
audience, and to provide a primarily statistical treatment of the
lessons that have been learned from this remarkable set of data.
\end{abstract}


\vspace*{-6pt}
\end{frontmatter}

\section{Introduction and Summary}  
\label{Section.Intro}

In what turned out to be an invaluable contribution to the research
community, Netflix Inc. of Los Gatos, California, on October 2, 2006,
publicly released a remarkable set of data, and offered a Grand Prize
of one million US dollars to the person or team who could succeed in
modeling this data to within a certain precisely defined predictive
specification. While this contest attracted attention from many
quarters---and most notably from within the computer science and
artificial intelligence communi\-ties---the heart of this contest was a
problem of statistical modeling, in a context known as \textit{collaborative filtering}. Our goal in this paper is to provide a
discussion and overview---from a primarily \textit{statistical}
viewpoint---of some of the lessons for statistics which emerged from
this contest and its data set. This vantage will also allow us to
search for alternative approaches for analyzing such data (while noting
some open problems), as well as to attempt to understand the
commonalities and interplay among the various methods that key
contestants have proposed.\looseness=1

Netflix, the world's largest internet-based movie rental company,
maintains a data base of ratings their users have assigned (from 1
``star'' to 5 ``stars'') to movies they have seen. The intended use of this
data is toward producing a system for recommending movies to users
based on predicting how much someone is going to like or dislike any
particular~mov\-ie. Such predictions can be carried out using information
on how much a user liked or disliked other movies they have rated,
together with information on how much other users liked or disliked
those same, as well as other, movies. Such \textit{recommender systems},
when sufficiently accurate, have considerable commercial value,
particularly in the context of the world wide web.

The precise specifications of the Netflix data are a bit involved, and
we postpone our description of it to Section \ref{Section.Data}. Briefly, however, the
\textit{training} data consists of some 100 million ratings made by
approximately 480,000 users, and involving some 18,000 movies. (The
corresponding ``matrix'' of user-by-\break movie ratings is thus almost 99\%
sparse.) A subset of about 1.5 million ratings of the training set,
called the \textit{probe} subset, was identified. A further data set,
called the \textit{qualifying} data was also supplied; it was divided
into two approximately equal halves, called the \textit{quiz} and \textit{test} subsets, each  consisting of about 1.5 million cases, but with
the ratings withheld. The probe, quiz and test sets were constructed to
have similar statistical properties.\footnote{Readers unfamiliar with
the Netflix contest may find it helpful to consult the more detailed
description of the data given in Section~\ref{Section.Data}.}

The Netflix contest was based on a root mean squared error (RMSE)
criterion applied to the three million predictions required for the
qualifying data. If one naively uses the overall average rating for
each movie on the training data (with the probe subset removed) to make
the predictions, then the RMSE attained is either 1.0104, 1.0528 or
1.0540, respectively, depending on whether it is evaluated in sample
(i.e., on the training set), on the probe set or on the quiz set.\vadjust{\goodbreak}
Netflix's own recommender system, called \textit{Cinematch}, which is
known to be based on computer-intensive but ``straightforward linear
statistical models with a lot of data conditioning'' is known to attain
(after fitting on the training data) an RMSE of either 0.9514 or
0.9525, on the quiz and test sets, respectively. (See Bennett and
Lanning, \citeyear{Bennett}.) These values represent, approximately, a
9$\frac{1}{2}$\% improvement over the naive movie-average predictor.
The contest's Grand Prize of one million US dollars was offered to
anyone who could first\footnote{Strictly, our use of ``first'' here is
slightly inaccurate owing to a Last Call rule of the competition.}
improve the predictions so as to attain an RMSE value of not more than
90\% of 0.9525, namely, 0.8572, or better, on the test set.

The Netflix contest began on Oct 2, 2006, and was to run until at least
Oct 2, 2011, or until the Grand Prize was awarded. More than 50,000
contestants internationally participated in this contest. Yearly
Progress Prizes of \$50,000 US were offered for the best improvement of
at least 1\% over the previous year's result. The Progress Prizes for
2007 and 2008 were won, respectively, by teams named ``BellKor'' and
``BellKor in BigChaos.'' Finally, on July 26, 2009, the Grand Prize
winning entry was submitted by the ``BellKor's Pragmatic Chaos'' team,
attaining RMSE values of 0.8554 and 0.8567 on the quiz and test sets,
respectively, with the latter value representing a 10.06\% improvement
over the contest's baseline. Twenty minutes after that submission (and
in accordance with the Last Call rules of the contest) a competing
submission was made by ``The Ensemble''---an amalgamation of many
teams---who attained an RMSE value of 0.8553 on the quiz set, and an RMSE
value of 0.8567 on the test set. To two additional decimal places, the
RMSE values attained on the test set were 0.856704 by the winners, and
0.856714 by the runners up. Since the contest rules were based on test
set RMSE, and also were limited to four decimal places, these two
submissions were in fact a tie. It is therefore the order in which
these submissions were received that determined the winner; following
the rules, the prize went to the earlier submission. Fearing legal
consequences, a second and related contest which Netflix had planned to
hold was canceled when it was pointed out by a~probabilistic argument
(see Narayanan and Shma\-tikov, \citeyear{Nar}) that, in spite of the precautions
taken to preserve anonymity, it might theoretically be possible to
identify some users on the basis of the seemingly limited information
in the data.

Of course, nonrepeatability, and other vagaries of ratings by humans,
itself imposes some lower bound on the accuracy that can be expected
from any recommender system, regardless of how ingenious it may be. It
now appears that the 10\% improvement Netflix required to win the
contest is close to the best that can be attained for this data. It
seems fair to say that Netflix technical staff possessed ``fortuitous
insight'' in setting the contest bar precisely where it did (i.e.,
0.8572 RMSE); they also were well aware that this goal, even if
attainable, would not be easy to achieve.

The Netflix contest has come and gone; in this story, significant
contributions were made by Yehuda Koren and by ``BellKor'' (R. Bell, Y.
Koren, C. Volinsky), ``BigChaos'' (M. Jahrer, A. Toscher), larger teams
called ``The Ensemble'' and ``Grand Prize,'' ``Gravity'' (G. Takacs, I.
Pilaszy, B. Nemeth,\break D.~Tikk), ``ML@UToronto'' (G.~Hinton, A.~Mnih, R.~Salakhutdinov;
``ML'' stands for ``machine learning''), lone contestant
Arkadiusz Paterek, ``Pragmat\-ic Theory'' (M.~Chabbert, M. Piotte) and many
others. Noteworthy of the contest was the craftiness of some
participants, and the open collaboration of others. Among such stories,
one that stands out is that of Brandyn Webb, a ``cybernetic
epistemologist'' having the alias Simon Funk (see Piatetsky, \citeyear{Piatetsky}). He
was the first to publicly reveal use of the SVD model together with a
simple algorithm for its implementation that allowed him to attain a
good early score in the contest (0.8914 on the quiz set). He also
maintains an engaging website at
\texttt{%
\href{http://sifter.org/\textasciitilde simon/journal}{http://}
\href{http://sifter.org/\textasciitilde simon/journal}{sifter.org/\textasciitilde simon/journal}}.

Although inspired by it, our emphasis in this paper is not on the
contest itself, but on the fundamentally different \textit{individual}
techniques which contribute to effective collaborative filtering
systems and, in particular, on the \textit{statistical} ideas which
underpin them. Thus, in Section \ref{Section.Data}, we first provide
a~careful description of the Netflix data, as well as a~number of
graphical displays. In Section \ref{Section.Notation} we establish the
notation we will use consistently throughout, and also include a~table
summarizing the performance of many of the methods discussed.
Sections~\ref{Section.ANOVA}, \ref{Section.SVD}, \ref{Section.RBM} and
\ref{Section.kNN} then focus on four key ``stand-alone'' techniques
applicable to the Netflix data. Specifically, in Section
\ref{Section.ANOVA} we discuss ANOVA techniques which provide a \textit{baseline} for most other methods. In Section \ref{Section.SVD} we
discuss the singular value decomposition or SVD (also known as the
latent factor model, or matrix factorization) which is arguably the
most effective single\vadjust{\goodbreak} procedure for collaborative filtering. A~fundamentally different paradigm is based on neural networks---in
particular, the restricted Boltzman machines (RBM)---which we describe
in Section~\ref{Section.RBM}. Last of these stand-alone methods are the
nearest neighbor or kNN methods which are the subject of Section~\ref{Section.kNN}.

Most of the methods that have been devised for collaborative filtering
involve parameterizations of very high dimension. Furthermore, many of
the models are based on subtle and substantive contextual insight. This
leads us, in Section \ref{Section.Shrink}, to undertake a~discussion of
the issues involved in dimension reduction, specifically penalization
and parameter\break shrinkage. In Section \ref{Section.Temporal} we digress
briefly to describe certain temporal issues that arise, but we return
to our main discussion in Section \ref{Section.New} where, after
exploring their comparative properties, and taking stock of the lessons
learned from the ANOVA, SVD, RBM and kNN models, we speculate on the
requisite characteristics of effective models as we search for new
model classes.

In response to the Netflix challenge, subtle, new and imaginative
models were proposed by many contest participants. A selection of those
ideas is summarized in Section~\ref{Section.Other}. At the end,
however, winning the actual contest proved not to be possible without
the use of many hybrid models, and without combining the results from
many prediction methods. This \textit{ensemble} aspect of combining many
procedures is discussed briefly in Section \ref{Section.Other}.
Significant computational issues are involved in a data set of this
magnitude; some numerical issues are described briefly in Section
\ref{Section.Numerical}. Finally, in Section \ref{Section.Conclusion},
we summarize some of the statistical lessons learned, and briefly note
a few open problems. In large part because of the Netflix contest, the
research literature on such problems is now sufficiently extensive that
a complete listing is not feasible; however, we do include a broadly
representative bibliography.
For earlier background and reviews, see, for example,
ACM SIGKDD (\citeyear{ACM}), Adomavicius and Tuzhilin (\citeyear{Adomavicius}),
Bell et al. (\citeyear{BellBennett}),
Hill et al. (\citeyear{Hill}), Hoffman (\citeyear{HofmannTwo}), Marlin (\citeyear{MarlinThree}),
Netflix (\citeyear{Netflix}), Park and Pennock (\citeyear{Park}), Pu et al. (\citeyear{Pu}),
Resnick and Varian (\citeyear{ResnickOne}) and Tuzhilin at al. (\citeyear{Tuzhilin}).

\section{The Netflix Data}              
\label{Section.Data}

In this section we provide a more detailed overview of the Netflix
data; these in fact consist of two key components, namely, a \textit{training}
set and a \textit{qualifying} set. The qualifying data set
itself consists of two halves,\vadjust{\goodbreak} called the \textit{quiz} set and the \textit{test}
set; furthermore, a particular subset of the training set,
called the \textit{probe} set, was identified. The quiz, test and probe
subsets were produced by a random three way split of a certain
collection of data, and so were intended to have identical statistical
properties.

The main component of the Netflix data---namely, the ``training'' set---can be
thought of as a matrix of ratings consisting of 480,189 rows,
corresponding to randomly selected anonymous users from Netflix's
customer base, and 17,770 columns, corresponding to movie titles. This
matrix is 98.8\% sparse; out of a possible $480{,}189 \times 17{,}770 =
8{,}532{,}958{,}530$ entries, only 100,480,507 ratings are actually available.
Each such rating is an integer value (a number of ``stars'') between 1
(worst) and 5 (best). The data were collected between October 1998 and
December 2005, and reflect the distribution of all ratings received by
Netflix during that period. It is known that Netflix's own database
consisted of over 1.9 billion ratings, on over 85,000 movies, from over
11.7 million subscribers; see Bennett and Lanning (\citeyear{Bennett}).

In addition to the training set, a qualifying data set consisting of
2,817,131 user--movie pairs was also provided, but with the ratings
withheld. It consists~of two halves: a quiz set, consisting of
1,408,342 user--movie pairs, and a test set, consisting of 1,408,789
pairs; these subsets were not identified. Contestants were required to
submit predicted ratings for the entire qualifying set. To provide
feedback to all participants, each time a contestant submitted a set of
predictions Netflix made public the RMSE value they attained on a
web-based \textit{Leaderboard}, but only for the quiz subset. Prizes,
however, were to be awarded on the basis of RMSE values attained on the
test~sub\-set. The purpose of this was to prevent contestants from tuning
their algorithms on the ``answer oracle.''

Netflix also provided the dates on which each of the ratings in the
data sets were made. The reason for this is that Netflix is more
interested in predicting future ratings than in explaining those of the
past. Consequently, the qualifying data set had been selected from
among the \textit{most recent} ratings that were made. However, to allow
contestants to understand the sampling characteristics of the
qualifying data set, Netflix identified the \textit{probe} subset of
1,408,395 user-movie pairs within the training set (and hence with
known ratings), whose distributional properties were meant to match
those of the qualifying data set. (The quiz, test and probe subsets
were produced from the random three-way split already mentioned.) As a
final point, prior to releasing their data, Netflix applied some\vadjust{\goodbreak}
statistically neutral perturbations (such as deletions, changes of
dates and/or ratings) to try to protect the confidentiality and
proprietary nature of its client base.

In our discussions, the term ``training set'' will generally refer to the
training data, but with the probe subset removed; this terminology is
in line with common usage when a subset is held out during a
statistical fitting process. Of course, for producing predictions to
submit to Netflix, contestants would normally retrain their algorithms
on the full training set (i.e., with probe subset included). As the
subject of our paper is more concerned with collaborative filtering
generally, rather than with the actual contest, we will make only
limited reference to the qualifying data set, and mainly in our
discussion on implicit data in Section \ref{Section.Other}, or when
indicating certain scores that contestants achieved.

Finally, we mention that Netflix also provided the titles, as well as
the release years, for all of the movies. In producing predictions for
its internal use, Netflix's Cinematch algorithm \textit{does} make use
of other data sources, such as (presumably) geographical, or other
information about its customers, and this allows it to achieve
substantial improvements in RMSE. However, it is known that Cinematch
does not use names of the movies, or dates of ratings. In any case, to
produce the RMSE values on which the contest was to be based, Cinematch
was trained without any such other data. Nevertheless, no restrictions
were placed on contestants from using external sources, as, for
instance, other databases pertaining to movies. Interestingly however,
none of the top contestants made use of any such auxiliary
information.\footnote{This is not to say they did not try. But perhaps
surpris\-ingly---with the possible exception of a more specific release
date---such auxiliary data did not improve RMSE. One possible
explanation for this is that the Netflix data set is large enough to
proxy such auxiliary information internally.}

\begin{figure}

\includegraphics{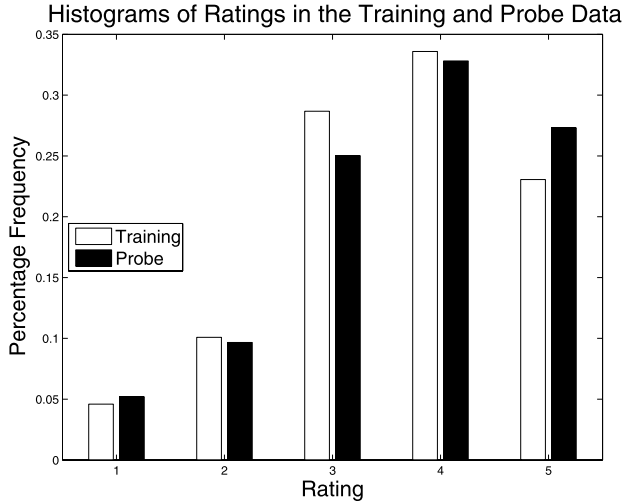}

\caption{Frequency histograms for ratings in the
training set (white) and probe set (black).}\label{f1}
\vspace*{3pt}
\end{figure}

\begin{figure}

\includegraphics{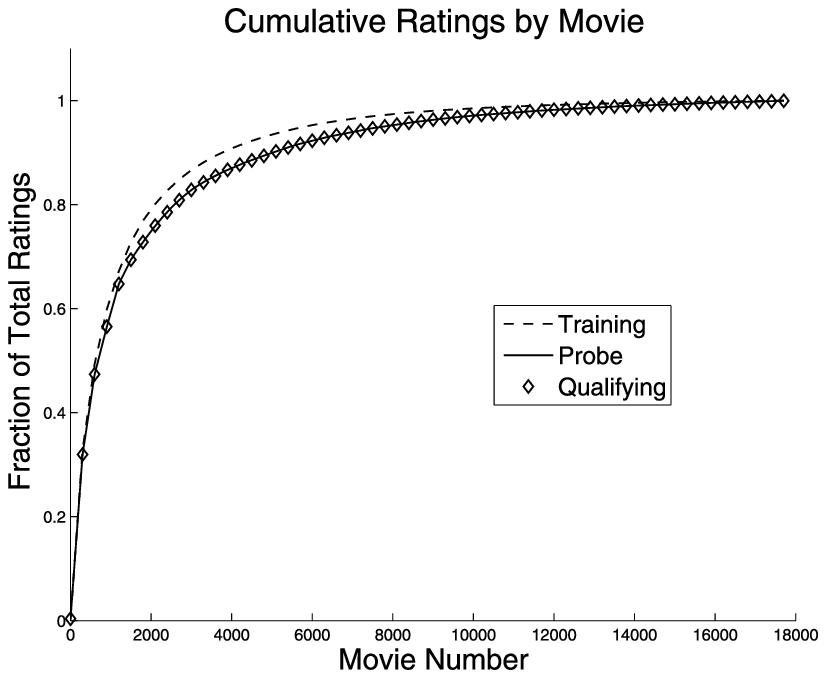}

\caption{Cumulative proportion of ratings, by movies,
for the training, probe and qualifying sets. Movies are on the
horizontal axis, sorted from most to least rated.}\label{f2}
\end{figure}

\begin{figure}[t]

\includegraphics{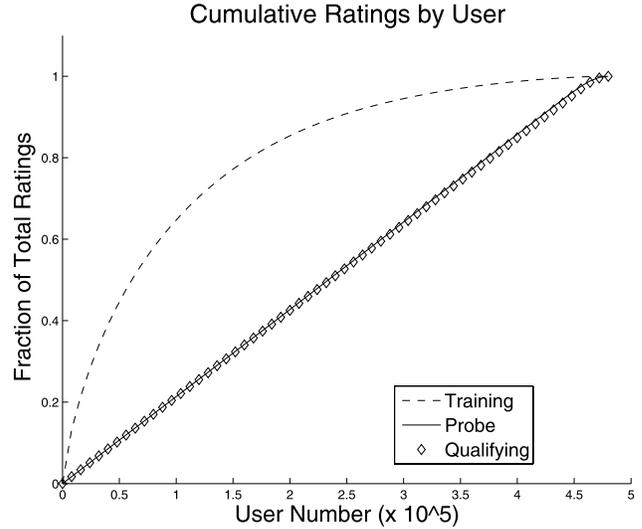}

\caption{Cumulative proportion of ratings, by users,
for the training, probe and qualifying sets. Users are on the
horizontal axis, sorted by number of movies rated (from most to
least).}\label{f3}
\end{figure}

\begin{figure}

\includegraphics{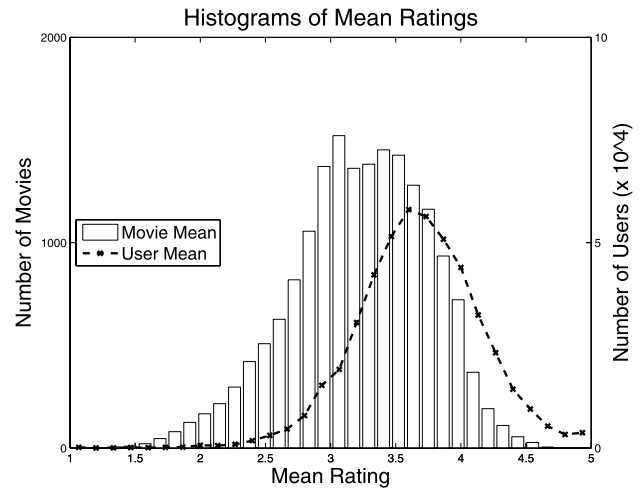}

\caption{Histograms for mean movie ratings (bars) and
mean user ratings (line with points) in the training set.}\label{f4}
\end{figure}

\begin{figure}

\includegraphics{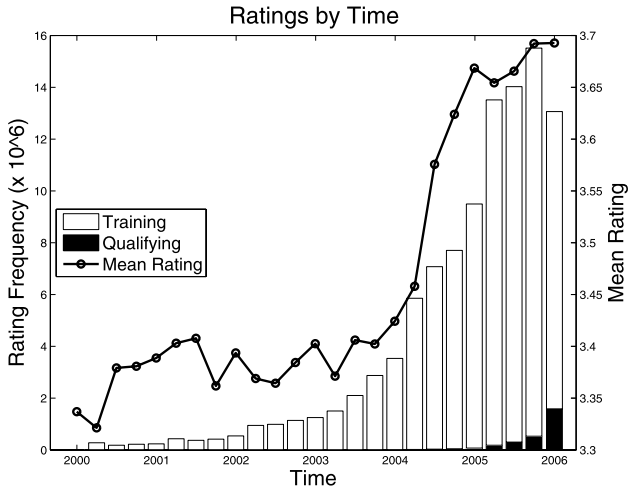}

\caption{Temporal effects: Histograms for the dates
(quarterly) of ratings in the training set (white bars) and qualifying
set (inlaid black bars). The line with points shows quarterly mean
ratings for the training data (scale on right).}\label{f5}\vspace*{-2pt}
\end{figure}

\begin{figure}

\includegraphics{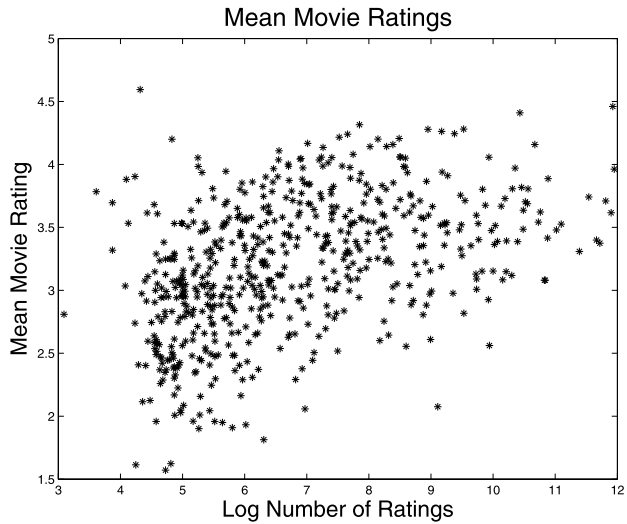}

\caption{Mean movie ratings versus (logarithm of)
number of ratings (for a random subset of movies).}\label{f6}\vspace*{5pt}
\end{figure}

Figures \ref{f1} through \ref{f6} provide some visualizations of the data. Figure \ref{f1}
gives histograms for the ratings in the training set, and in the probe
set. Reflecting temporal effects to be discussed in Section~\ref{Section.Temporal}
(but see also Figure~\ref{f5}), the overall mean
rating, 3.6736, of the probe set is significantly higher than the
overall mean, 3.6033, of the training set. Figures \ref{f2} and~\ref{f3} are plots
(in lieu of histograms) of the cumulative number of ratings in the
training, probe and qualifying sets. Figure \ref{f2} is cumulative by movies
(horizontal axis, and sorted from most to least rated in the training
set), while Figure~\ref{f3}\vadjust{\goodbreak} is cumulative by users. The steep rise in Figure \ref{f2}
indicates, for instance, that the 100 and 1000 most rated movies
account for over 14.3\% and 62.5\% of the ratings, respectively. In
fact, the most rated\footnote{We remark here that rented movies can be
rated without having been watched.} movie (Miss Congeniality) was rated
by almost half the users in the training set, while the least rated was
rated only 3 times. Figure \ref{f2} also evidences a slight---although
statistically significant---difference in the profiles for the
training and the qualifying (and probe) data. In Figure~\ref{f3}, the
considerable mismatch between the curve for the training data, with the
curves for the probe and qualifying sets which match closely, reflects
the fact that the representation of user viewership in the training set
is markedly different from that of the cases for which predictions are
required; clearly, Netflix constructed the qualifying data to have a
much more uniform distribution of user viewership.

Figure \ref{f4} provides histograms for the movie mean ratings and the user
mean ratings. The reason for the evident mismatch between these two
histograms is that the best rated movies were watched by
disproportionately large numbers of users. Finally, Figure \ref{f5}
exemplifies some noteworthy temporal effects in the data. Histograms
are shown for the number of ratings, quarterly, in the training and in
the qualifying data sets, with the dates in the qualifying set being
much later than the dates in the training set. Figure \ref{f5} also shows a
graph of the quarterly mean ratings for the training data (with the
scale being at the right). Clearly, a significant rise in mean rating
occurred starting around the beginning of 2004. Whether this occurred
due to changes in the ratings definitions, to the introduction of a
recommender system, to a change in customer profile, or due to some
other reason, is not known.

Finally, Figure \ref{f6} plots the mean movie ratings against the (log) number
of ratings. (Only a random sample of movies is used so as not to
clutter the plot.) The more rated movies do tend to have the higher
mean ratings but with some notable exceptions, particularly among the
less rated movies which can sometimes have very high mean ratings.

As a general comment, the layout of the Netflix data contains enormous
variation. While the average number of ratings per user is 209, and the
average number of ratings per movie is 5654.5 (over the entire training
set), five users rated over 10,000 movies each, while many rated fewer
than 5 movies. Likewise, while some movies were rated tens of thousands
of times, most were rated fewer than 1000 times, and many less than 200
times. The extent of this variation\vadjust{\goodbreak} implies large differences in the
accuracies with which user and movie parameters can be estimated, a
problem which is particularly severe for the users. Such extreme
variation is among the features of typical collaborative filtering data
which complicate their analyses.

\section{Notation and a Summary Table}         
\label{Section.Notation}

In this section we establish the notation we will adhere to in our
discussions throughout this paper. We also include a table, which will
be referred to in the sequel, of RMSE performance for many of the
fitting methods we will discuss.

\begin{table*}
\tablewidth=15cm
\caption{RMSE values attained by various methods}\label{t1}
\begin{tabular*}{15cm}{@{\extracolsep{\fill}}lcl@{}}
\hline
\textbf{Predictive model}      & \textbf{RMSE}      & \textbf{Remarks and references}   \\
\hline
$\hat r_{i,j} = \mu$            &  1.1296  &  RMSE on probe set, using mean of training set \\
$\hat r_{i,j} = \alpha_i$       &  1.0688  &  Predict by user's training mean, on probe set \\
$\hat r_{i,j} = \beta_j$        &  1.0528  &  Predict by movie's training mean, on probe set \\
$\hat r_{i,j} = \mu + \alpha_i + \beta_j$, naive  &  0.9945  &  Two-way ANOVA, no iteraction \\
$\hat r_{i,j} = \mu + \alpha_i + \beta_j$         &  0.9841  &  Two-way ANOVA, no iteraction \\
``Global effects''                &  0.9657  &  Bell and Koren (\citeyear{BellThree,BellSix,BellFive}) \\
Cinematch, on quiz set          &  0.9514  &  As reported by Netflix \\
Cinematch, on test set          &  0.9525  &  Target is to beat this by 10\% \\
kNN                        &  0.9174  &  Bell and Koren (\citeyear{BellThree,BellSix,BellFive}) \\
``Global''${}+{}$SVD             &  0.9167  &  Bell and Koren (\citeyear{BellThree,BellSix,BellFive}) \\
SVD                        &  0.9167  &  Bell and Koren (\citeyear{BellThree,BellSix,BellFive}), on probe set \\
``Global''${}+{}$SVD${}+{}$``joint kNN''   &  0.9071  &  Bell and Koren (\citeyear{BellThree,BellSix,BellFive}), on probe set \\
``Global''${}+{}$SVD${}+{}$``joint kNN''   &  0.8982  &  Bell and Koren (\citeyear{BellThree,BellSix,BellFive}), on quiz set \\
Simon Funk                 &  0.8914  &  An early submission; Leaderboard \\
TemporalDynamics${}+{}$SVD$++$   &  0.8799  &  Koren (\citeyear{KorenThree}) \\
Arkadiusz Paterek's best score        &  0.8789  &  An ensemble of many methods; Leaderboard \\
ML Team: RBM${}+{}$SVD         &  0.8787  &  See Section \ref{Section.RBM}; Leaderboard  \\
Gravity's best score       &  0.8743  &  November 2007; Leaderboard \\
Progress Prize, 2007, quiz    &  0.8712  &  Bell, Koren and Volinsky (\citeyear{BellFour,BellOne,Bell}) \\
Progress Prize, 2007, test    &  0.8723  &  As above, but on the test set \\
Progress Prize, 2008, quiz    &  0.8616  &  Bell, Koren and Volinsky (\citeyear{BellTwo}), Toscher and Jahrer (\citeyear{Toscher}) \\
Progress Prize, 2008, test    &  0.8627  &  As above, but on the test set \\
Grand Prize, target        &  0.8572  &  10 \% below Cinematch's RMSE on test set \\
Grand Prize, runner up     &  0.8553  &  The Ensemble, 20 minutes too late; on quiz set \\
Grand Prize, runner up     &  0.8567  &  As above, but on the test set \\
Grand Prize, winner        &  0.8554  &  BellKor${}+{}$BigChaos${}+{}$PragmaticTheory, on quiz set \\
Grand Prize, winner        &  0.8567  &  As above, but on the test set
\\
\hline
\end{tabular*}
\legend{Selected RMSE values, compiled from various sources. Except as
noted, RMSE values shown are either for the probe set after fitting on
the training data with the probe set held out, or for the quiz set
(typically from the Netflix Leaderboard) after fitting on the training
data with the probe set included.}\vspace*{-3pt}
\end{table*}

Turning to notation, we will let $i = 1, 2, \ldots , I$ range over the
users (or their indices) and $j = 1, 2, \ldots , J $ range over the
movies (or their indices). For the Netflix training data, $I =
480{,}189$ and $J = 17{,}770$. Next, we will let $J(i)$ be the set of
movies rated by user $i$ and $I(j)$ be the set of users who rated
movie $j$. The cardinalities of these sets will be denoted variously as
$J_i \equiv |J(i)|$ and $I_j \equiv |I(j)|$. We shall also use the
notation $\mathcal C$ for the set of all user-movie pairs $(i,j)$ whose
ratings are given. Denoting the total number of user-movie ratings in
the training set by $N$, note that $N = |{\mathcal C}| = \sum_{i=1}^I J_i =
\sum_{j=1}^J I_j$. The ratings are made on an ordered scale (such
scales are known as ``Likert scales'') and are coded as integers having
values $k = 1, 2, \ldots , K$; for Netflix, $K=5$. The actual ratings
themselves, for $(i,j) \in \mathcal C$, will be denoted by~$r_{i,j}$.
Averages of $r_{i,j}$ over $i \in I(j)$, over $j \in J(i)$, or over
$\mathcal C$ (i.e., over movies, or users, or over the entire training data
set) will be denoted by~$r_{\bolds\cdot ,j}$, $r_{i, \bolds\cdot}$\vadjust{\goodbreak} and
$r_{\bolds\cdot, \bolds\cdot}$ respectively. Estimated values are denoted by
``hats'' as in $\hat r_{i,j}$, which may refer to the fitted value from a
model when $(i,j) \in \mathcal C$, or to a~predicted value otherwise. Many
of the procedures we discuss are typically fitted to the residuals from
a baseline fit such as an ANOVA; where this causes no confusion, we
continue using the notations $r_{i,j}$ and $\hat r_{i,j}$ in referring
to such residuals. Some procedures, however, involve both the original
ratings as well as their residuals from other fits; in such cases, the
residuals are denoted as $e_{i,j}$ and $\hat e_{i,j}$. Finally, the
notation $I(j, j')$ will refer to the set of all users who saw both
movies $j$ and $j'$, and $J(i,i')$ will refer to the set of movies that
were seen by both users $i$ and~$i'$.\looseness=-1

%
%

Finally, we also include, in this section, a table which provides a
summary, compiled from multiple sources, of the RMSE values attained by
many of the methods discussed in this paper. The RMSE values shown in
Table \ref{t1} are typically for the probe set, after fitting on the remainder
of the training set; or where known,\vadjust{\goodbreak} on the quiz set, after fitting on
the entire training set; but exceptions to this are noted. References
to the ``Leaderboard'' refer to performance on the quiz set publicly
released by Neflix. We refer to Table \ref{t1} in our subsequent discussions.

\section{ANOVA Baselines}      
\label{Section.ANOVA}

ANOVA methods furnish baselines for many analyses. One basic
approach---referred to as preprocess\-ing---involves first removing \textit{global} effects such as user and movie means, and using the residuals
as input to subsequent models. Alternatively, such ``row'' and ``column''
effects can be incorporated directly into those models where they are
sometimes referred to as biases. In any case, most models work best
when global effects are explicitly accounted for. In this section we
discuss minimizing the sum of squared errors criterion
\begin{equation}\label{MSE.Criterion}
\operatorname{\sum\sum}\limits_{(i,j) \in \mathcal C}  ( r_{i,j} - \hat r_{i,j}  )^2\vadjust{\goodbreak}
\end{equation}
using various ANOVA methods for the predictions $\hat r_{i,j}$ of the
user-movie ratings $r_{i,j}$. Due to the large number of parameters,
regularization (i.e., penalization) would normally be used, but we
reserve our discussions of regularization issues to Section~\ref{Section.Shrink}.

We first note that the best fitting model of the form
\begin{equation}\label{fit.const}
\hat r_{i,j} \equiv \mu \quad   \mbox{ for all } i, j,
\end{equation}
obtained by setting $\mu = $ 3.6033, the mean of all user-movie ratings
in the training set (with probe removed), results in an RMSE on the
training set equal to its standard deviation 1.0846; on the probe set,
using this same $\mu$ results in an RMSE of 1.1296, although the actual
mean and standard deviation for the probe set\footnote{Note that the
difference between the squares of the probe's 1.1296 and 1.1274 RMSE
values must equal the squared difference between the two means, 3.6736
and 3.6033.} are 3.6736 and 1.1274.

Next, if we predict each rating by the mean rating for that user on the
training set, thus fitting the model
\begin{equation}\label{fit.users}
\hat r_{i,j} = \mu + \alpha_i  ,
\end{equation}
we obtain an RMSE of 0.9923 on the training set, and 1.0688 using the
same values on the probe. If, instead, we predict each rating by the
mean for that movie, thus fitting
\begin{equation}\label{fit.movies}
\hat r_{i,j} = \mu + \beta_j  ,
\end{equation}
we obtain RMSE values 1.0104 and 1.0528 on the training and probe sets,
respectively. The solutions for (\ref{fit.const})--(\ref{fit.movies})
are just the least squares fits associated with
\begin{eqnarray}\label{LSE.criteria}
&&\operatorname{\sum \sum}\limits_{(i,j) \in \mathcal C}  ( r_{i,j} - \mu   )^2 ,\nonumber
\\
&&\operatorname{\sum
\sum}\limits_{(i,j) \in \mathcal C}  ( r_{i,j} - \mu - \alpha_i   )^2  \quad\mbox{and}
\\
&&\operatorname{\sum \sum}\limits_{(i,j) \in \mathcal C}  ( r_{i,j} - \mu - \beta_j\nonumber
 )^2 ,
\end{eqnarray}
respectively, where $\mathcal C$ is the set of indices $(i,j)$ over the
training set. Histograms of the user and movie means were given in
Figure \ref{f4}; we note, for later use, that the the variances of the user
and movie means on the test set (with probe removed) are 0.23074 and
0.27630, corresponding to standard deviations of 0.48035 and 0.52564,
respectively.\footnote{These values are useful for assessing
regularization issues; see Section \ref{Section.Shrink}.}

We now consider two-factor models of the form
\begin{equation}\label{fit.users.movies}
\hat r_{i,j} = \mu + \alpha_i + \beta_j  .
\end{equation}
Identifiability conditions, such as $\sum_i \alpha_i = 0$ and\break $\sum_j
\beta_j = 0$, would normally be imposed, although they become
unnecessary under typical regularization. If we were to proceed as in a
balanced two-way layout (i.e., with no ratings missing), then we would
first estimate $\mu$ as the mean of all available ratings; the values
of $\alpha_i$ and $\beta_j$ would then be estimated as the row and
column means, over the available ratings, after $\mu$ has been
subtracted throughout. Doing this results in RMSE values of 0.9244 and
0.9945 on the training and probe sets. If we proceed sequentially, the
order of the operations for estimating the~$\alpha_i$'s and the
$\beta_j$'s will matter: If we estimate the~$\alpha_i$'s first and
subtract their effects before estimating the~$\beta_j$'s, the result
will not be the same as first estimating the~$\beta_j$'s and
subtracting their effect before estimating the $\alpha_i$'s; these
procedures result, respectively, in RMSE values of 0.9218 and 0.9177 on
the training set.

The layout for the Netflix data is unbalanced, with the vast majority
of user-movie pairings not rated; we therefore seek to minimize
\begin{equation}\label{Criterion.mab}
\sum_{(i,j) \in \mathcal C}  ( r_{i,j} -\mu -\alpha_i -\beta_j )^2
\end{equation}
over the training set. This quadratic criterion is convex, however,
standard methods for solving the ``normal'' equations, obtained by
setting derivatives with respect to $\mu$, $\alpha_i$ and $\beta_j$ to
zero, involve matrix inversions which are not feasible over such high
dimensions. The optimization of (\ref{Criterion.mab}) may, however, be
carried out using either an EM or a gradient descent algorithm. When no
penalization is imposed, minimizing (\ref{Criterion.mab}) results in an
RMSE value of 0.9161 on the training set, and 0.9841 on the probe
subset.

A consideration when fitting (\ref{fit.users.movies}) as well as other
models is that some predicted ratings $\hat r_{i,j}$ can fall outside
the $[1,5]$ range. This can occur when highly rated movies are rated by
users prone to giving high ratings, or when poorly rated movies are
rated by users prone to giving low ratings. Under optimization of
(\ref{Criterion.mab}) over the test set, approximately 5.1 million
$\hat r_{i,j}$ estimates fall below 1, and 19.4 million fall above 5.
Although we may Winsorize (clip) these~$\hat r_{i,j}$ to lie in
$[1,5]$, clipping in advance need not be optimal when residuals from a
baseline fit are input to other procedures. We do not consider here the
problem of minimizing (\ref{Criterion.mab}) when $\mu + \alpha_i +
\beta_j$ there is replaced by a Winsorized version.\vadjust{\goodbreak}

Of course, not all is well here. The differences in RMSE values between
the training and the probe sets reflect temporal effects, some of which
were already noted. Furthermore, these models have parameterizations of
high-dimensions and have therefore been overfit, resulting in inferior
predictions. These issues will be dealt with in Sections
\ref{Section.Shrink} and \ref{Section.Temporal}.

Finally, we remark that interaction terms can be added to
(\ref{fit.users.movies}). The standard approach $\hat r_{i,j} = \mu +
\alpha_i + \beta_j + \gamma_{i,j}$ will not be effective, although it
could possibly be combined with regularization. Alternatively,
interactions could be based on user $\times$ movie groupings via ``many
to one'' functions $a(i)$ and $b(j)$, and models such as
\begin{equation}\label{fit.groups}
\hat r_{i,j} = \mu + \alpha_i + \beta_j + \gamma_{a(i),b(j)}  .
\end{equation}
There are many possibilities for defining such groups; for example, the
covariates discussed in Section \ref{Section.Other} or nearest neighbor
methods (kNN) can be used to construct suitable $a(i)$ and $b(j)$. Some
further interaction-type ANOVA models are considered in Section~\ref{Section.New}.

\section{SVD Methods}          
\label{Section.SVD}

In statistics, the singular value decomposition\break (SVD) is best known for
its connection to principal components: If $X = (X_1 , X_2 , \ldots ,
X_n)^\prime$ is a random vector of means $0$, and $n \times n$
covariance matrix $\Sigma$, then one may represent $\Sigma$ as a linear
combination of mutually orthogonal rank 1 matrices, as in
\[
\Sigma = \sum_{j=1}^n \lambda_j P_j P_j^\prime  ,
\]
where $\lambda_1 \geq \lambda_2 \geq \cdots \geq \lambda_n \geq 0$ are
ordered eigenvalues of $\Sigma$, and $P_j$ corresponding orthonormal
(column) eigenvectors. The principal components are the random
variables $P_j^\prime X$. Less commonly known is that the $n \times n$
matrix\vspace*{1pt} $ {\mathcal T}^{(k)} = \sum_{j=1}^k \lambda_j P_j P_j^\prime$ gives
the best rank $k$ reconstruction of $\Sigma$, in the sense of
minimizing the Frobenius norm $ \| \Sigma - {\mathcal T}^{(k)}\|$, defined
as the square root of the sum of the squares of its entries.

These results generalize. If $A$ is an arbitrary real-valued $m \times
n$ matrix, its singular value decomposition is given by\footnote{If
$A$ is complex-valued, these relations still hold, with conjugate
transposes replacing transposes.}
\[
A = U D V^\prime,
\]
where $U\,{=}\,(U_1 , U_2 , \ldots , U_m)$ is an $m \times m$ matrix whose
columns $U_j$ are orthonormal eigenvectors of $A A^\prime$, where $V =
(V_1 , V_2 , \ldots , V_n)$\vadjust{\goodbreak} is an $n \times n$ matrix whose columns
$V_j$ are orthonormal eigenvectors of $A^\prime A$, and where $D$ is an
$m \times n$ ``diagonal'' matrix whose diagonal entries  may be taken as
the descending order nonnegative values
\begin{eqnarray*}
\lambda_j    &=&   + \sqrt {  \{ \operatorname{eigval}  A A^\prime \}_j }
\\
&=&
 +
\sqrt {  \{\operatorname{eigval}  A^\prime A  \}_j }  ,\quad   j = 1 , 2, \ldots ,
\min (m,n),
\end{eqnarray*}
called the singular values of $A$. The columns of $V$ and $U$ provide
natural bases for inputs to and outputs from the linear transformation
$A$. In particular, $A V_j = \lambda_j U_j$, $A^\prime U_j = \lambda_j
V_j$, so given an input $B = \sum_{j=1}^n c_j V_j$, the corresponding
output is $AB = \sum_{j=1}^{{\rm min} (m,n) } \lambda_j c_j U_j$.\vspace*{2pt}

Given an SVD of $A$, the Eckart--Young Theorem states that, for a given
$k < \min (m,n)$, the best rank~$k$ reconstruction of $A$, in the sense
of minimizing the Frobenius norm of the difference, is $U^{(k)} D^{(k)}
 ( V^{(k)}  )^\prime$, where $U^{(k)}$ is the $m \times k$ matrix
formed from the first~$k$ columns of $U$, $V^{(k)}$ is the $n \times k$
matrix formed by the first $k$ columns of $V$, and $D^{(k)}$ is the
upper left $k \times k$ block of $D$. This reconstruction may be
expressed in the form $FG^\prime$ where $F$ is $m \times k$ and $G$ is
$k \times n$; the reconstruction is thus formed from the inner products
between the $k$-vectors comprising $F$ with those comprising $G$. These
$k$-vectors may be thought of as associated, respectively, with the
rows and the columns of~$A$, and (in applications) the components of
these vectors are often referred to as \textit{features}. A numerical
consequence of the Eckart--Young Theorem is that ``best'' rank $k$
approximations can be determined iteratively: given a best rank $k-1$
approximation, $FG^\prime$, say, a best rank $k$ approximation is
obtained by attaching a column vector to each of~$F$ and $G$ which
provide a best fit to the residual matrix $A - FG^\prime$. SVD
algorithms can therefore be quite straightforward. Here, however, we
are specifically concerned with algorithms applicable to matrices which
are sparse. We briefly discuss two such algorithms, useful in
collaborative filtering, namely, alternating least squares (ALS) and
gradient descent. Some relevant references are
Bell and Koren (\citeyear{BellFive}),
Bell, Koren and Volinsky (\citeyear{BellFour}),
Funk (\citeyear{Funk}),
Koren, Bell and Volinsky (\citeyear{KorenTwenty}),
Raiko, Ilin and Karhunen (\citeyear{Raiko}),
Srebro and Jaakkola (\citeyear{Srebro}),
Srebro, Rennie and Jaakkola (\citeyear{SrebroTwo}),
Takacs et al. (\citeyear{TakacsTwo,TakacsOne,TakacsThree,TakacsFour}),
Wu (\citeyear{Wu}) and
Zhou et al. (\citeyear{Zhou}).
See also
Hofmann (\citeyear{HofmannThree},
\citeyear{Hofmann}),
Hofmann and Puzicha (\citeyear{HofmannFour}),
Kim and Yum (\citeyear{Kim}),
Marlin and Zemel (\citeyear{MarlinTwo}),
Rennie and Srebro (\citeyear{Rennie}),
Sali (\citeyear{Sali}) and
Zou et al. (\citeyear{Zou}).\vadjust{\goodbreak}

The alternating least squares (ALS) method for determining the best
rank $p$ reconstruction involves expressing the summation in the
objective function in two ways:
\begin{eqnarray}\label{three.ways}
&&\sum\sum_{\mathcal C}  \Biggl( r_{i,j}  -  \sum_{k=1}^p u_{i,k} v_{j,k}
 \Biggr)^2\nonumber
 \\
  &&\quad= \sum_{j=1}^J  \sum_{i \in I(j)}  \Biggl( r_{i,j} -
\sum_{k=1}^p u_{i,k} v_{j,k}  \Biggr)^2
\\
&&\quad = \sum_{i=1}^I  \sum_{j \in J(i)}
 \Biggl( r_{i,j} -  \sum_{k=1}^p u_{i,k} v_{j,k}  \Biggr)^2 .\nonumber
\end{eqnarray}
The $u_{i,k}$ may be initialized using small independent normal
variables, say. Then, for each fixed $j$, we carry out the least
squares fit for the $v_{j,k}$ based on the inner sum in the middle
expression of (\ref{three.ways}). And then, for each fixed $i$, we
carry out the least squares fit for $u_{i,k}$ based on the inner sum of
the last expressions in (\ref{three.ways}). This procedure is iterated
until convergence; several dozen iterations typically
suffice.

ALS for SVD with regularization\footnote{\label{footnote.lambda}Although we prefer to postpone discussion of regularization to the
unified treatment attempted in Section \ref{Section.Shrink}, it is
convenient to lay out those SVD equations here.} proceeds similarly.
For example, minimizing\footnote{We prefer not to set $\lambda_1 =
\lambda_2$ at the outset for reasons of conceptual clarity; see Section
\ref{Section.Shrink}. In fact, because a constant may pass freely
between user and movie features, generality is not lost by taking
$\lambda_1 = \lambda_2$. Generality is lost, however, when these values
are held constant across all features; see Section~\ref{Section.Shrink}.}
\begin{eqnarray}\label{SVD.ALS.regular}
&&\sum\sum_{\mathcal C}  \Biggl( r_{i,j}  -  \sum_{k=1}^p u_{i,k} v_{j,k}
 \Biggr)^2 \nonumber
 \\[-8pt]\\[-8pt]
 &&{}+ \lambda_1  \sum_{i=1}^I \|u_i\|^2 + \lambda_2
\sum_{j=1}^J \| v_j \|^2\nonumber
\end{eqnarray}
leads to iterations which alternate between minimizing
\begin{equation}\label{SVD.ALS.regular.one}
\sum_{i \in I(j)}  \Biggl( r_{i,j} -  \sum_{k=1}^p u_{i,k} v_{j,k}
 \Biggr)^2
+ \lambda_1 \|v_j\|^2
\end{equation}
with respect to the $v_{j,k}$, and then minimizing
\begin{equation}\label{SVD.ALS.regular.two}
\sum_{j \in J(i)}  \Biggl( r_{i,j} -  \sum_{k=1}^p u_{i,k} v_{j,k}
 \Biggr)^2
+ \lambda_2 \|u_i\|^2
\end{equation}
with respect to the $u_{i,k}$; these are just ridge regression
problems.\footnote{Some contestants preferred the regularization
\[
\sum\sum_{\mathcal C}  \Biggl[    \Biggl( r_{i,j}  -  \sum_{k=1}^p u_{i,k} v_{j,k} \Biggr)^2
+ \lambda  ( \|u_i\|^2 + \|v_j\|^2   )  \Biggr]
\]
instead of (\ref{SVD.ALS.regular}), which changes the $\lambda_1$ and
$\lambda_2$ in (\ref{SVD.ALS.regular.one}) and~(\ref{SVD.ALS.regular.two}) into $I_j \lambda$ and $J_i \lambda$,
respectively. In Section \ref{Section.Shrink} we argue that this
modification is not theoretically optimal.}

ALS can also be performed one feature at a time, with the advantage of
yielding factors in descending order of importance. To do this, we
initialize as before, and again arrange the order of summation in the
objective function in two different ways; for the first feature, this
is
\begin{eqnarray}\label{two.ways}
&&\sum_{j=1}^J  \biggl[ \sum_{i \in I(j)}  ( r_{i,j} - u_{i,1}v_{j,1}
 )^2  \biggr] \nonumber
 \\[-8pt]\\[-8pt]
 &&\quad=  \sum_{i=1}^I  \biggl[ \sum_{j \in J(i)}  (
r_{i,j} - u_{i,1}v_{j,1}  )^2  \biggr]  .\nonumber
\end{eqnarray}
We then iterate between the least squares problems of the inner sums in
(\ref{two.ways}), namely,
\begin{equation}\label{two.ways.A}
\hat v_{j,1} = \sum_{i \in I(j)} u_{i,1} r_{i,j}  \Big/ \sum_{i \in I(j)}
u_{i,1}^2
\end{equation}
for all $j$, and then
\begin{equation}\label{two.ways.B}
\hat u_{i,1} = \sum_{j \in J(i)} v_{j,1} r_{i,j}  \Big/ \sum_{j \in J(i)}
v_{j,1}^2
\end{equation}
for all $i$, until convergence. After $k-1$ features have been fit, we
compute the residuals
\[
e_{i,j} = r_{i,j} - \sum_{\ell=1}^{k-1} u_{i,\ell}v_{j,\ell}
\]
and replace (\ref{two.ways.A}) and (\ref{two.ways.B}) by
\[
\hat v_{j,k} = \sum_{i \in I(j)} u_{i,k} e_{i,j}  \Big/ \sum_{i \in I(j)}
u_{i,k}^2
\]
and
\[
    \hat u_{i,k} = \sum_{j \in J(i)}
v_{j,k} e_{i,j}  \Big/ \sum_{j \in J(i)} v_{j,k}^2   ,
\]
ranging over all $j$ and all $i$, respectively.

Regularization in one-feature-at-a-time ALS can be effected in several
ways. Bell, Koren and Volinsky (\citeyear{BellFour}) shrink the residuals $e_{i,j}$
via
\[
e_{i,j} \leftarrow \frac{n_{i,j}}{n_{i,j} + \lambda_k} e_{i,j}  ,
\]
where $n_{i,j} = \min (I_j, J_i)$ measures the ``support''\break for~$r_{i,j}$,
and they increase the shrinkage parameter~$\lambda_k$ with each feature
$k$. Alternately, one could add a~regularization term
\[
\lambda_k  ( \| u_k\|^2 + \| v_k\|^2   )
\]
when fitting the $k$th feature, choosing the $\lambda_k$ by
cross-validation.

Finally, we consider gradient descent approaches for fitting SVD
models. For an SVD of dimension $p$, say, we first initialize all
$u_{i,k}$ and $v_{j,k}$ in
\[
\sum\sum_{\mathcal C}  \Biggl( r_{i,j} - \sum_{k=1}^p u_{i,k} v_{j,k}
 \Biggr)^2 .
\]
Then write
\[
e_{i,j} = r_{i,j} - \sum_{k=1}^p u_{i,k} v_{j,k}  ,
\]
and note that
\[
\frac{\partial e_{i,j}^2} {\partial u_{i,k}} = - 2 e_{i,j} v_{j,k}
\]
 and
 \[
  \frac{\partial e_{i,j}^2} {\partial v_{j,k}} = - 2
e_{i,j} u_{i,k}  .
\]
Updating can then be done locally following the negative gradients:
\begin{eqnarray}\label{Grad.Update.One}
u_{i,k}^{\mathrm{new}} &=& u_{i,k}^{\mathrm{old}}  + 2 \eta e_{i,j} v_{j,k}^{\mathrm{old}}  \quad
\mbox{and} \nonumber
\\[-8pt]\\[-8pt]
   v_{j,k}^{\mathrm{new}} &=& v_{j,k}^{\mathrm{old}} + 2 \eta
e_{i,j} u_{i,k}^{\mathrm{old}}  ,\nonumber
\end{eqnarray}
where the \textit{learning rate} $\eta$ controls for overshoot. For a
given $(i,j) \in \mathcal C$, these equations are used to update the
$u_{i,k}$ and $v_{j,k}$ for all $k$; we then cycle over the $(i,j) \in
\mathcal C$ until convergence. If we regularize\footnote{If, instead of
(\ref{Grad.Update.better}), we regularized as
\[
\sum\sum_{\mathcal C}  [  ( r_{i,j} - u_i^\prime v_j   )^2 + \lambda (
\|u_i\|^2 + \|v_j\|^2  )  ],
\]
then the gradient descent update equations (\ref{Grad.Update.Three})
become
\begin{eqnarray*}
u_{i,k}^{\mathrm{new}} &= &u_{i,k}^{\mathrm {old}}  + \eta  ( 2 e_{i,j} v_{j,k}^{\mathrm
{old}} - \lambda u_{i,k}^{\mathrm {old}}   )    \quad \mbox{and}
\\
 v_{j,k}^{\mathrm {new}} &=&
v_{j,k}^{\mathrm {old}}  + \eta  ( 2 e_{i,j} u_{i,k}^{\mathrm {old}} - \lambda
v_{j,k}^{\mathrm {old}}   )  .
\end{eqnarray*}\vspace*{-15pt}
} the problem, as in
\begin{eqnarray}\label{Grad.Update.better}
&&\sum\sum_{\mathcal C}  \Biggl( r_{i,j} - \sum_{k=1}^p u_{i,k} v_{j,k}
 \Biggr)^2\nonumber
 \\[-8pt]\\[-8pt]
&&\quad{}+ \lambda   \biggl( \sum_i \|u_i\|^2 + \sum_j \|v_j\|^2  \biggr)  ,\nonumber
\end{eqnarray}
the update equations become
\begin{eqnarray}\label{Grad.Update.Three}
\hspace*{6pt}\qquad u_{i,k}^{\mathrm{new}} &=& u_{i,k}^{\mathrm{old}}  + \eta  \biggl( 2 e_{i,j} v_{j,k}^{\mathrm{old}} - \frac{\lambda}{J_i} u_{i,k}^{\mathrm{old}}   \biggr)\quad\mbox{and} \nonumber
\\[-8pt]\\[-8pt]
v_{j,k}^{\mathrm{new}}& =& v_{j,k}^{\mathrm{old}}  + \eta  \biggl( 2 e_{i,j} u_{i,k}^{\mathrm{old}} -
\frac{\lambda}{I_j} v_{j,k}^{\mathrm{old}}  \biggr)  .\nonumber
\end{eqnarray}
We note that, as in ALS, there are other ways to sequence the updating
steps in gradient descent. Simon Funk (\citeyear{Funk}), for instance, trained the
features one at a time. To train the $k$th feature, one
initializes the $u_{i,k}$ and $v_{j,k}$ randomly, and then loops over
all \mbox{$(i,j) \in {\mathcal C}$}, updating the $k$th feature for all users and
all movies. The updating equations are as before [e.g.,
(\ref{Grad.Update.Three})]
except based on residuals $e_{i,j} = r_{i,j} - \sum_{\ell = 1}^k u_{i,
\ell} v_{j, \ell}$. After convergence, one proceeds to the next
feature.

We remark that sparse SVD problems are known to be nonconvex and to
have multiple local minima; see, for example, Srebro and Jaakkola
(\citeyear{Srebro}). Nevertheless, starting from different initial conditions, we
found that SVD seldom settled into entirely unsatisfactory minima,
although the minima attained did vary slightly. The magnitude of these
differences was commensurate with the variation inherent among the
options available for regularization. We also found that averaging the
results from several SVD fits started at different initial conditions
could lead to better results than a single SVD fit of a higher
dimension. On this point, see also Wu (\citeyear{Wu}). Finally, we note the
recent surge of work on a problem referred to as matrix completion;
see, for example, Candes and Plan (\citeyear{Candestwo}).

\section{Neural Networks and RBMs}     
\label{Section.RBM}

\begin{figure}

\includegraphics{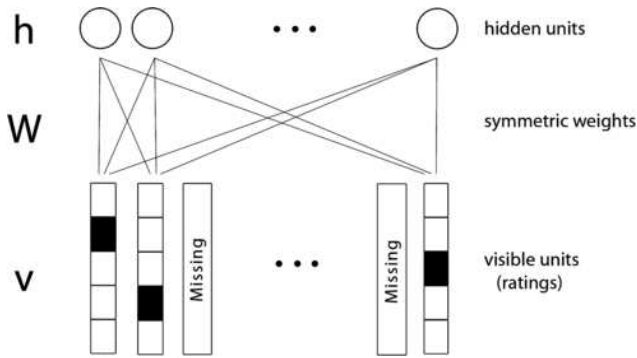}

\caption{The RBM model for a single user: Each of the
user's hidden units is connected to every visible unit (a multinomial
observation) that represents a rating made by that user. Every user is
associated with one RBM, and the RBM models for the different users are
linked through the common symmetric weight parameters
$W_{j,f}^k$.}\label{f7}
\end{figure}

A restricted Boltzman machine (RBM) is a neural network consisting of
one layer of \textit{visible} units, and one layer of \textit{invisible}
ones; there are no \textit{connections} between units \textit{within}
either of these layers, but all units of one layer are connected to all
units of the other layer. To be an RBM, these connections must be
bidirectional and symmetric; some definitions require that the units
only take on binary values, but this restriction is unnecessary. We
remark that the symmetry condition is only needed so as to simplify the
training process. See Figure~\ref{f7}; additional clarification will emerge
from the discussion below. The name for these networks derives from the
fact that their governing probability distributions are analogous to
the Boltzman distributions which arise in statistical mechanics. For
further background, see, for example, Hertz, Krogh and Palmer (\citeyear{Hertz}), Section
7.1, Izenman (\citeyear{Izenman}), Chapter 10,
or Ripley (\citeyear{Ripley}), Section~8.4. See also Bishop (\citeyear{BishopTwo}, \citeyear{Bishop}). We will describe the RBM model that has
been applied to the Netflix data by Salakhutdinov, Mnih and
  Hinton (\citeyear{Salakhutdinov}), whom we
will also refer to as SMH.


In the SMH model, to each user $i$, there corresponds a length $F$
vector of hidden (i.e., unobserved) units, or features, $h =  ( h_1 ,
h_2 , \ldots , h_F  )$. These features, $h_f$, for $f = 1, 2, \ldots ,
F$, are random variables posited to take on binary values, 0 or 1. Note
that subscripting to indicate the dependence of $h$ on the $i$th
user has been suppressed. Next, instead of thinking of the ratings of
the $i$th user as the collection of values $r_{i,j}$ for $j \in
J(i)$, we think of this user's ratings as the collection of vectors
$v_j =  ( v_j^1 , v_j^2 , \ldots , v_j^K  )$, for $j \in J(i)$, that
is,\vspace*{1.5pt} for each of the movies he or she has seen. Each of these vectors is
defined by setting all of its elements to 0, except for one: namely,
$v_j^k = 1$, corresponding to $r_{i,j} = k$. Here $K$ is the number of
possible ratings; for Netflix, $K=5$. The collection of these $v_j$
vectors for our $i$th user [with $j \in J(i)$] will be denoted
by $v$. Here again, the dependence of~$v$, as well as of the~$v_j$ and
the $v_j^k$, on the user $i$ is suppressed.

We next introduce the symmetric weight parameters $W_{j,f}^k$ for $1
\leq j \leq J$, $1 \leq f \leq F$\vspace*{1pt} and $1 \leq k \leq K$, which link
each of the $F$ hidden features of a user with each of the $J$ possible
movies; these weights also carry a dependence on the rating values
$k$.\break
The~$W_{j,f}^k$ are not dependent on the user; the same weights apply
to all users, however, only weights for the movies he or she has rated
will be relevant for any particular user.

We next specify the underlying stochastic model. First, the
distributions of the $(v, h)$ are assumed to be independent across
users. We therefore only need to specify a probability distribution on
the collection $(v, h)$ for the $i$th user. This distribution is
determined by its two conditional distributions modeled as follows: The
conditional distribution of the $i$th user's observed ratings
information $v$, given that user's hidden features vector $h$, is
modeled as a (one-trial) multinomial distribution
\begin{eqnarray}\label{Hinton.One}
&&P  (   v_j^k = 1   |    h    )   \nonumber
\\[-8pt]\\[-8pt]
&&\quad =   \frac{\exp  ( b_j^k +
\sum_{f=1}^F h_f W_{j,f}^k  )} {\sum_{\ell = 1}^{K} \exp  ( b_j^\ell +
\sum_{f=1}^F h_f W_{j,f}^\ell  ) }  ,\nonumber
\end{eqnarray}
where the denominator is just a normalizing factor. Next, the
conditional distributions of the $i$th user's hidden features
variables, given that user's observed ratings $v$, are modeled as
conditionally independent Bernoulli variables
\begin{equation}\label{Hinton.Two}
\qquad P  (  h_f = 1  |   v    )    =   \sigma  \Biggl( b_f + \sum_{j \in J(i)}
\sum_{k=1}^K v_j^k W_{j,f}^k  \Biggr)  ,
\end{equation}
where $\sigma(x) = 1 /  (1 + e^{-x }  )$ is the sigmoidal function.
Note that (\ref{Hinton.Two}) is equivalent to the linear logit model
\begin{eqnarray}\label{Hinton.Two.B}
&&\log  \biggl(   \frac {P  (  h_f = 1   |  v   )} {1 - P  ( h_f = 1 |  v )}
\biggr)   \nonumber
\\[-8pt]\\[-8pt]
&&\quad=   b_f + \sum_{j \in J(i)} \sum_{k=1}^K v_j^k W_{j,f}^k  ;\nonumber
\end{eqnarray}
in effect, (\ref{Hinton.Two})/(\ref{Hinton.Two.B}) models user features
in terms of the movies the user has rated, and the user's ratings for
them. Note that the weights (interaction parameters) $W_{j,f}^k$ are
assumed to act symmetrically in (\ref{Hinton.One}) and
(\ref{Hinton.Two}). The parameters $b_j^k$ and $b_f$ are referred to as
biases; the $b_j^k$ may be initialized to the logs of their respective
sample proportions over all users. We remark that in this model there
is no analogue for user biases.

To obtain the joint density of $v$ and $h$ from their two conditional
distributions, we make use of the following result: Suppose $f(x,y)$ is
a joint density for $(X,Y)$, and that $f_1(x|y)$, $f_2(y|x)$ are the
corresponding conditional density functions for $X|Y$ and $Y|X$. Then
noting the elementary equalities
\begin{eqnarray*}
f(x,y)   &=&   f_1(x|y) \times \frac{f_2(y|x^*)}{f_1(x^* | y)} \times
f_X(x^*)
\\
 &=&   f_2(y|x) \times \frac{f_1(x|y^*)}{f_2(y^* | x)} \times
f_Y(y^*)  ,
\end{eqnarray*}
we see that $f(x,y)$ can be determined from $f_1$ and~$f_2$ since it is
proportional to either of
\[
f_1(x|y) \times \frac{f_2(y|x^*)}{f_1(x^* | y)}   \quad  \mbox{and}\quad f_2(y|x)
\times \frac{f_1(x|y^*)}{f_2(y^* | x)}  .
\]
Here $f_X$ and $f_Y$ are the marginals of $X$ and $Y$, and the choices
of $x^*$ and $y^*$ are arbitrary. It follows that the joint density of
$(v,h)$ satisfies the proportionality
\[
p(v,h)   \propto   \frac{P_2(h|v) P_1(v|h^*)}{P_2(h^* |v)}  ;
\]
with the choice $h^* = 0$, this yields
\[
p(v,h)   \propto   \exp \{ - E(v,h) \}  ,
\]
where
\begin{eqnarray*}
E(v,h) &=& -\sum_{j \in J(i)} \sum_{f=1}^F \sum_{k=1}^K W_{j,f}^k h_f
v_j^k - \sum_{j \in J(i)} \sum_{k=1}^K v_j^k b_j^k
\\
&&{} - \sum_{f=1}^F h_f
b_f + \sum_{j \in J(i)} \log  \Biggl( \sum_{k =1}^K b_j^k  \Biggr)  .
\end{eqnarray*}
The computations here just involve taking products over the observed
ratings using (\ref{Hinton.One}), and over the hidden features using
(\ref{Hinton.Two}). By analogy to formulae in statistical mechanics,
$E(v,h)$ is referred to as an \textit{energy}; note that only movies
whose ratings are known contribute to it. The joint density of $(v,h)$
can therefore be expressed as
\[
p(v,h) = \frac{\exp \{ - E(v,h) \}} { \sum_{v', h'} \exp \{ - E(v',h')
\} }  ,
\]
so that the likelihood function (i.e., the marginal distribution for
the observed data) is
\begin{equation}\label{Hinton.like}
p(v) = \frac{ \sum_h \exp \{ - E(v,h) \}} { \sum_{v', h'} \exp \{ -
E(v',h') \} }  .
\end{equation}
We will use the notation
\[
Z = \sum_{v'} \sum_{h'} \exp  \bigl( - E(v',h')  \bigr)
\]
for the denominator term of (\ref{Hinton.like}).

Now the updating protocol for the $W_{j,f}^k$ is given by
\[
\Delta W_{j,f}^k \equiv \varepsilon  \frac {\partial \log p(v)} {\partial
W_{j,f}^k}  ,
\]
where $\varepsilon$ is a ``learning rate.'' To determine $\Delta W_{ij}^k$,
we will need the derivatives
\[
\frac {\partial E(v,h)} {\partial W_{j,f}^k} =  - h_f v_j^k
\]
and
\[
\frac {\partial Z} {\partial W_{j,f}^k} = \sum_{v'} \sum_{h'} \exp
 \{ -  E(v',h')  \}  h_f^\prime {v_j^k}^\prime  .
\]
Now
\begin{eqnarray}\label{yuhe}
\frac{{\partial \log  p(v)}}{{\partial W_{j,f}^k}} &=& \frac{{\partial
\log  ( \sum_h {\exp ( - E(v,h))}  )}} {{\partial W_{j,f}^k}} \nonumber
\\[-8pt]\\[-8pt]
&&{}-
\frac{{\partial \log Z}}{{\partial W_{j,f}^k}}  ;\nonumber
\end{eqnarray}
the first term on the right in (\ref{yuhe}) equals
\begin{eqnarray*}
&&\frac{1}  { \sum_h \exp ( - E(v,h)) } \sum_h  \exp \bigl( - E(v,h)\bigr) h_f
v_j^k
\\
&&\quad= \sum_h p(h|v){h_j}v_i^k  ,
\end{eqnarray*}
while the second term on the right in (\ref{yuhe}) is
\begin{eqnarray*}
\frac{1}{Z} \frac{ \partial Z } { \partial W_{j,f}^k } &=& \frac{1}{Z}
\sum_{v'} {\sum_{h'} \exp  \bigl( - E(v',h')  \bigr) } h_f' {v_j^k}^\prime
\\
& =&
\sum_{v'} \sum_{h'} p  ( v',h'  ) h_f' {v_j^k}^\prime  .
\end{eqnarray*}
Hence, altogether,
\begin{eqnarray*}
&&\Delta W_{j,f}^k
\\
&&\quad= \varepsilon  \biggl( \sum_h p(h|v) h_f v_j^k  - \sum_{v'}
\sum_{h'} p  ( v',h'  ) h_j' {v_i^k}^\prime  \biggr) ,
\end{eqnarray*}
or, expressed more concisely,
\begin{equation}\label{SMH.update.one}
\Delta W_{j,f}^k = \varepsilon  ( \langle v_j^k h_f \rangle _{\mathrm{data}} - \langle v_j^k h_f
\rangle _{\mathrm{model}} )  .
\end{equation}
Similarly, we obtain the updating protocols
\begin{equation}\label{SMH.update.two}
\Delta b_f = \varepsilon  ( \langle h_f \rangle _{\mathrm{data}} - \langle h_f \rangle _{\mathrm{model}} )
\end{equation}
and
\begin{equation}\label{SMH.update.three}
\Delta b_j^k = \varepsilon  ( \langle v_j^k \rangle _{ \mathrm{data}} - \langle v_j^k \rangle _{\mathrm{model}} )  .
\end{equation}
Note that the gradients here are for a single user only; therefore, the
three updating equations (\ref{SMH.update.one}), (\ref{SMH.update.two})
and (\ref{SMH.update.three}) must first be averaged over all users.

The updating equations (\ref{SMH.update.one}), (\ref{SMH.update.two})
and (\ref{SMH.update.three}) for implementing maximum likelihood
``learning'' in-\break volves two forms of averaging. The averaging over the
``data,'' that is, based on the $p(h|v)$, is relatively straightforward.
However, the averaging over the ``model'' is impractical, as it requires
Gibbs-type MCMC sampling from $p(v,h)$ which involves iterating between
(\ref{Hinton.One}) and (\ref{Hinton.Two}). SMH instead suggest running
this Gibbs sampler for only a small number of steps at each stage, a
procedure referred to as ``contrastive divergence'' (Hinton, \citeyear{Hinton}). For
further details, we refer the reader to SMH.

Numerous variations on the model defined by (\ref{Hinton.One}) and
(\ref{Hinton.Two}) are possible. In particular, the user features $h$
may be modeled as Gaussian variables having, say, unit variances. In
this case the model for $P(  v_j^k = 1  |  h  )$ remains as at
(\ref{Hinton.One}), but (\ref{Hinton.Two}) becomes
\begin{eqnarray*}
&&P  (  h_f = h  |  v   )
\\
&&\quad=   \frac{1}{\sqrt{2 \pi}} \exp  \Biggl\{ -
\frac{1}{2}  \Biggl( h - b_f - \sum_{j \in J(i)} \sum_{k=1}^K v_j^k W_{j,
f}^k  \Biggr)^2  \Biggr\}  .
\end{eqnarray*}
The marginal distribution $p(v)$ remains as at (\ref{Hinton.like})
except with energy term
\begin{eqnarray*}
E(v,h) &=& -\sum_{j \in J(i)} \sum_{f=1}^F \sum_{k=1}^K W_{j,f}^k h_f
v_j^k - \sum_{j \in J(i)} \sum_{k=1}^K v_j^k b_j^k
\\
&&{}+ \frac{1}{2}
\sum_{f=1}^F  ( h_f - b_f  )^2 + \sum_{j \in J(i)} \log  \Biggl( \sum_{k =
1}^K b_j^k  \Biggr)  .
\end{eqnarray*}
The parameter updating equations remain unchang\-ed. Salakhutdinov, Mnih and
Hinton (\citeyear{Salakhutdinov}) report that this Gaussian version does
not perform as well as
the binary one; perhaps the nonlinear structure in (\ref{Hinton.Two})
is useful for modeling the Netflix data. Bell, Koren and Volinsky
(\citeyear{BellOne,BellTwo}), on the other hand, prefer the Gaussian model.

SMH indicate that to contrast sufficiently among users, good models
typically require the number of binary user features to be not less
than about $F=100$. Hence, the dimension of the weights $W$, which is
$J \times F \times K$, can be upward of ten million. The
parameterization of $W$ can be reduced somewhat by representing it as a
product of matrices of lower rank, as in $W_{j,f}^k = \sum_{\ell = 1}^p
A_{j \ell}^k B_{\ell f}$. This approach reduces the number of $W$
parameters to $J \times p \times K + p \times F$, a~factor of about
$p/F$.

There is a further point which we mention only briefly here, but return
to in Section \ref{Section.Other}. While the Netflix qualifying data
omits ratings, it does provide \textit{implicit} information in the form
of \textit{which} movies users chose to rate; this is particularly
useful for users having only a small number of ratings in the training
set. In fact, the full binary matrix indicating which user-movie pairs
were rated (regardless of whether or not the ratings are known) is an
important information source. This information is valuable because the
values missing in the ratings matrix are not ``missing at random'' and
for purposes of the contest, exploiting this information was critical.
It turns out that RBM models can incorporate such implicit information
in a relatively straightforward way; according to Bell, Koren and
Volinsky (\citeyear{BellOne}), this is a key strength of RBM models. For further
details we refer the reader to SMH.

SMH reported that, when they also incorporated this implicit
information, RBMs slightly outperform\-ed carefully-tuned SVD models.
They also found that the errors made by these two types of models were
significantly different so that linearly combining multiple RBM and SVD
models, using coefficients determined over the probe set, allowed them
to achieve an error rate over 6\% better than Cinematch. The
ML@UToronto team's Leaderboard score ultimately attained an RMSE of
0.8787 on the quiz set (see Table \ref{t1}).

\section{Nearest Neighbor (kNN) Methods}              
\label{Section.kNN}

Early recommender systems were based on nearest neighbors (kNN)
methods, and have the advantage of conceptual and computational
simplicity useful for producing convincing explanations to users as to
why particular recommendations are being made to them. Usually applied
to the residuals from a preliminary fit, kNN tries to identify
(pairwise) similarities among users or among movies and use these to
make predictions. Although generally less accurate than SVD, kNN models
capture local aspects of the data not fitted completely by SVD or other
global models we have described. Key references include Bell and Koren
(\citeyear{BellThree,BellSix,BellFive}),
Bell, Koren and Volinsky (\citeyear{BellFour,BellTwo}),  
Koren (\citeyear{KorenOne,KorenTwo}),                          
Sarwar et al. (\citeyear{SarwarTwo}), Toscher, Jahrer and
  Legenstein (\citeyear{Toschertwo}) and Wang, de~Vries and Reinders (\citeyear{Wang}).
See also Herlocker et al. (\citeyear{HerlockerOne}), Tintarev and Masthoff (\citeyear{Tintarev})
and Ungar and Foster (\citeyear{Ungar}).

While the kNN paradigm applies symmetrically to movies and to users, we
focus our discussion on movie nearest neighbors, as these are the more
accurately estimable, however, both effects are actually important. A
basic kNN idea is to estimate the rating that user $i$ would assign to
movie $j$ by means of a weighted average of the ratings he or she has
assigned to movies most similar to $j$ among movies which that user has
rated:
\begin{equation}\label{basic.kNN}
\hat r_{i,j} = \frac{ \sum_{j' \in N(j;i)}  s_{j,j'}  r_{i,j'} } {
\sum_{j' \in N(j;i)} s_{j,j'} }  .
\end{equation}
Here the $s_{j,j'}$ are similarity measures which act as weights, and
$N(j;i)$ is the set of, say, $K$ movies, that~$i$ has seen and that are
most similar to $j$. Letting $I(j,j') = I(j) \cap I(j')$ be the set of
users who have seen both movies $j$ and $j'$, similarity between pairs
of movies can be measured using Pearson's correlation
{\fontsize{10.3}{12.3}{\selectfont{\[
s_{j,j'} = \frac{  \sum_{i \in I(j,j')}  ( r_{i,j} - r_{\bolds\cdot, j}  )
 ( r_{i,j'} - r_{ \bolds\cdot, j'}  )   }
 {
    \sqrt { \sum_{i \in
I(j,j')}  ( r_{i,j}  - r_{\bolds\cdot, j}  )^2  }
 \sqrt { \sum_{i \in
I(j,j')}  ( r_{i,j'} - r_{ \bolds\cdot, j'}  )^2 }  } ,
\]}}}%
or by the variant
{\fontsize{10.3}{12.3}{\selectfont{\[
s_{j,j'} =  \frac{\sum_{i \in I(j,j')}  ( r_{i,j} - r_{i, \bolds\cdot}  )
 ( r_{i,j'} - r_{i, \bolds\cdot}  )}
 { \sqrt { \sum_{i \in I(j,j')}
 ( r_{i,j} - r_{i,\bolds\cdot}  )^2   }
\sqrt { \sum_{i \in I(j,j')}
 ( r_{i,j'} - r_{i, \bolds\cdot}  )^2 } }
\]}}}%
in which centering is at the user instead of the movie means, or by
cosine similarity
\[
s_{j,j'} = \frac{  \sum_{i \in I(j,j')}  r_{i,j}  r_{i,j'}  } {   \sqrt
{  \sum_{i \in I(j,j')} r_{i,j}^2  }
 \sqrt {  \sum_{i \in I(j,j')} r_{i,j'}^2 }   }  .
\]
The similarity measure is used to determine the nearest neighbors, as
well as to provide the weights\break in~(\ref{basic.kNN}). In practice, if an
ANOVA, SVD and/or other fit is carried out first, kNN would be applied
to the residuals from that fit; under such ``centering'' the behavior of
the three similarity measures above would be very alike. As the common
supports $I(j,j')$ vary greatly, it is usual to regularize the
$s_{j,j'}$ via a~rule such as
\[
s_{j,j'}    \leftarrow   \frac{ |I(j,j')| } { |I(j,j')| +  \lambda }
s_{j,j'}  .
\]

A more data-responsive kNN procedure could be based on
\[
\hat r_{i,j}  = \sum_{j' \in N(j;i)} w_{j,j'} r_{i,j'}  ,
\]
where the weights $w_{j,j'}$ (which are specific to the $i$th user) are
meant to be chosen via least squares fits
\begin{equation}\label{kNN.LS}\qquad
\arg \min_w {\sum_{i' \ne i}  \biggl( r_{i',j}  - \sum_{j' \in N(j;i)}
w_{j,j'} r_{i',j'}  \biggr)}^2  .
\end{equation}
This procedure cannot be implemented effectively as shown because
enough $r_{i',j'}$ ratings are often not available, however, Bell and
Koren (\citeyear{BellFive}) suggest how one may compensate for the missing ratings
here in a natural way.

Many variations of such methods can be proposed and can produce
slightly better estimates, although at an increased computational
burden; see Bell, Koren and Volinsky (\citeyear{BellTwo}) and Koren
 (\citeyear{KorenOne,KorenTwo}). For
example, user-specific weights, with their relatively\vadjust{\goodbreak} inaccurate local
optimizations, could be replaced by global weights having a relatively
more accurate global optimization, as in the model
\begin{eqnarray}\label{Koren.big.kNN}
\hat r_{i,j} &=& b_{i,j} + \sum_{j^\prime \in N^k(j;i)}  ( r_{i,
j^\prime} -  b_{i, j^\prime}    ) w_{j, j^\prime} \nonumber
\\[-8pt]\\[-8pt]
&& {}+ \sum_{j^\prime \in
N^k(j;i)} C_{i, j^\prime}  .\nonumber
\end{eqnarray}
Here $w_{j, j^\prime}$ is the same for all users, and the neighborhoods
are now $N^k(j;i) \equiv J(i) \cap N^k(j)$, where~$N^k(j)$ is the set
of $k$ movies most similar to $j$ as determined by the similarity
measure. The sum involving the $C_{i, j^\prime}$ is included in order
to model implicit information inherent in the choice of movies a user
rated; for purposes of the Netflix contest, this sum would include the
cases in the qualifying data. As a~further enhancement, the $b_{i,j}$
following the equality and the $b_{i, j^\prime}$ within the sum could
be decoupled, with the second of these remaining as the original
baseline values, and the first of these set to $\mu + a_i + b_j$ and
then trained simultaneously with the model. Furthermore, the sums in
(\ref{Koren.big.kNN}) could each be normalized, for instance, using
coefficients such as $ | N^k(j;i)  |^{-1/2}$.

\section{Dimensionality and Parameter Shrinkage}               
\label{Section.Shrink}

The large number (often millions) of parameters in the models discussed
make them prone to overfitting, affecting the accuracy of the
prediction process. Reducing dimensionality through penalization
therefore becomes mission critical. This leads to considerations which
are relatively recent in statistics, such as the effective number of
degrees of freedom of a regularized model and its use in assessing
predictive accuracy, as well as to the connections between that
viewpoint and James--Stein shrinkage and empirical Bayes ideas. In this
section we attempt to place such issues within the Netflix context. A
difficulty which arises here stems from the distributional mismatch
between the training and validation data, however, we will sidestep
this issue so as to focus on key theoretical considerations. Our
discussion draws from Casella (\citeyear{Cassella}),
Copas (\citeyear{Copas}),   
Efron (\citeyear{EfronONE,EfronTWO,EfronTHREE,EfronFOUR,EfronFIVE}), Efron et al. (\citeyear{EfronEtAl}), Efron
and
Morris (\citeyear{EfronMorrisOne,EfronMorrisTwo,EfronMorrisThree,EfronMorrisFour,EfronMorrisFive,EfronMorrisSix,EfronMorrisSeven}), Houwelingen (\citeyear{Houwelingen}),
Morris (\citeyear{Morris}), Stein (\citeyear{SteinTwo}), Ye (\citeyear{y1998}) and Zou et al.
(\citeyear{z2007}). See also Barbieri and Berger (\citeyear{Barbieri}), Baron (\citeyear{Baron}),
Berger (\citeyear{Berger}), Breiman and Friedman (\citeyear{BreimanFried}),
Candes and Tao (\citeyear{Candes}),\vadjust{\goodbreak} Carlin and Louis (\citeyear{Carlin}),
Fan and Li (\citeyear{Fan}), Friedman (\citeyear{Friedman}),
Greenshtein and Ritov\break (\citeyear{Greenshtein}), Li (\citeyear{Li}),
Mallows (\citeyear{Mallows}), Maritz and Lwin (\citeyear{Maritz}), Moody (\citeyear{Moody}),
Robins (\citeyear{Robbins}, \citeyear{RobbinsTwo}, \citeyear{RobbinsThree}), Sarwar et al. (\citeyear{SarwarOne}),
Stone (\citeyear{Stone}) and Yuan and Lin (\citeyear{Yuan}).

\subsection*{Prediction Optimism}

To give some context to our
discussion, suppose~$Y$ is an $n \times 1$ random vector with entries
$Y_i$, for $i = 1, 2, \ldots , n$, all having finite second moment, and
suppose the mean of $Y$ is modeled by a vector $\mu(\beta)$ with
entries $\mu_i(\beta)$, where $\beta$ is a $p \times 1$ vector of
parameters. We will assume $\mu(\beta)$ is twice differentiable, and
that it uniquely identifies $\beta$. We also assume that there is a
unique value of $\beta$, namely, $\beta_0$, for which $Y$ can be
modeled as
\begin{equation}
Y_i = \mu_i(\beta_0) + e_i  , \label{Y.mu.model}
\end{equation}
with the $e_i$ then assumed to have zero means, equal variances $\operatorname{Var}( e_i  ) = \sigma^2$, and to be uncorrelated. The vector
$\mu(\beta)$ may or may not be based on a known, fixed design matrix
$X$; all that matters about $X$ is that it is considered known, and
that it fully determines the stochastic properties of $Y$.

Now let $Y^*$, with entries $Y_i^*$, be a stochastically independent
copy of $Y$ also defined on $X$, that is, on the same experiment. We
consider expectation~$E$ to be defined on the joint probability
structure of $(Y,Y^*)$ or, more precisely, of $(Y,Y^*) | X$; sometimes~$E$ will act on a function of $Y$ alone, and sometimes on a function of
both $Y$ and $Y^*$. Our starting point is the pair of inequalities
\begin{eqnarray}\label{Inequality}
 E \inf_\beta \sum_{i=1}^n  [  Y_i - \mu_i(\beta)   ]^2   &<& \inf_\beta E
\sum_{i=1}^n  [  Y_i - \mu_i(\beta)   ]^2 \hspace*{-30pt} \nonumber
\\[-8pt]\\[-8pt]
&<&  E \sum_{i=1}^n  [ Y_i^* -
\mu_i(\hat \beta)  ]^2 ,\nonumber\hspace*{-30pt}
\end{eqnarray}
which clearly will be strict, except in degenerate situations. The
infimum in the middle expression is assumed to occur at the value
$\beta = \beta_0$ identified at~(\ref{Y.mu.model}). The infimum inside
the expectation on the left occurs at the value of $\beta$ denoted as
$\hat \beta$; we\vspace*{1pt} interchangeably use the notation $\hat \beta^{(n)}$,
$\hat \beta (Y)$ and $\hat \beta^{(n)} (Y)$ for $\hat \beta$ when we
wish to stress its dependence on the sample size $n$, on the data~$Y$,
or on both. The $\mu_i (\hat \beta)$ occurring in the rightmost
expression in~(\ref{Inequality}) refers to entries of $\mu ( \hat
\beta^{(n)} (Y) )$, so that the $Y_i^*$ and $\mu_i (\hat \beta)$ there
are independent. The inequalities (\ref{Inequality}) have the
interpretation
\begin{equation}\label{Inequality.two}
E (\mbox{training error})\,{<}\,n \sigma^2\,{<}\,E(\mbox{prediction
error}) ,\hspace*{-35pt}
\end{equation}
it being understood that here the predictions $\mu(\hat \beta)$ are for
an independent repetition $Y^*$ of the same random experiment. Efron
(\citeyear{EfronTWO}) refers to the difference between prediction error and fitted
error, that is, between the right- and left-hand sides in
(\ref{Inequality})/(\ref{Inequality.two}), as the \textit{optimism}.

It is helpful, for the sake of exposition, to examine the inequalities
(\ref{Inequality})/(\ref{Inequality.two}) for a linear model, where
$\mu(\beta) = X \beta$, and $X$ is $n \times p$. In that case, the
leftmost and rightmost expressions in (\ref{Inequality}) are
equidistant from the middle one, and
(\ref{Inequality})/(\ref{Inequality.two}) become
\begin{equation}\label{Inequality.three}
(n - p) \sigma^2    <    n \sigma^2    <    (n+p) \sigma^2  .
\end{equation}
Here the leftmost evaluation follows from the standard regression
ANOVA, and corresponds to the fact that unbiased estimation of
$\sigma^2$ requires dividing the training error sum of squares by
$n-p$, while the rightmost evaluation follows from
\begin{eqnarray*}
E \sum_{i=1}^n  [  Y_i^* - \mu_i(\hat \beta)  ]^2 &=& E \sum_{i=1}^n
 [ \mu_i(\beta_0) + e_i^* - \mu_i(\hat \beta)  ]^2
 \\
 &=& n \sigma^2 +
E \sum_{i=1}^n  [  \mu_i(\beta_0) - \mu_i(\hat \beta)  ]^2 ,
\end{eqnarray*}
where the last expectation here evaluates as
\begin{eqnarray*}
&&E ( X \beta_0 - X \hat \beta )^\prime ( X \beta_0 - X \hat \beta )
\\
&&\quad= E
(\beta_0 - \hat \beta)^\prime (X^\prime X) (\beta_0 - \hat \beta)
\\
&&\quad = p
\sigma^2
\end{eqnarray*}
since $\beta_0 - \hat \beta$ has mean $0$ and covariance $\sigma^2
(X^\prime X)^{-1}$.

The inequalities (\ref{Inequality})/(\ref{Inequality.two}) hold whether
or not we have a linear model $\mu(\beta) = X \beta$, but the exact
evaluations of their left- and right-most terms as at
(\ref{Inequality.three}) do not. However, these evaluations (as well as
their equidistances from $n \sigma^2$) continue to hold asymptotically:
if the dimension of $\beta$ stays fixed at~$p$, and if the design $X$
changes with $n$ in such a way that the convergence of the least
squares estimate~$\hat \beta^{(n)}$ to $\beta_0$ is
$\sqrt{n}$-consistent, then both
\begin{equation}\label{Equality.one}
\qquad\lim_{n \rightarrow \infty}  \Biggl\{  n \sigma^2 - E \inf_\beta \sum_{i=1}^n
 [  Y_i - \mu_i(\beta)   ]^2  \Biggr\}   =   p \sigma^2
\end{equation}
and
\begin{equation}\label{Equality.two}
\qquad\lim_{n \rightarrow \infty}   \Biggl\{ E \sum_{i=1}^n  [  Y_i^* - \mu_i(\hat
\beta)  ]^2 - n \sigma^2  \Biggr\}   =    p \sigma^2
 .
\end{equation}
The proofs involve Taylor expanding $\mu(\hat \beta^{(n)})$ around
$\beta = \beta_0$ (recall $\mu$ is twice differentiable) and following
the proofs for the linear case; terms in the expansion beyond the
linear one are inconsequential by the $\sqrt{n}$-consistency.

\subsection*{Effective Degrees of Freedom}

The distances
$p \sigma^2$ across both boundaries in (\ref{Inequality.three}), as
well as at (\ref{Equality.one}) and (\ref{Equality.two}), lead to a
natural definition for the \textit{effective number of degrees of
freedom} of a statistical fitting procedure. In the linear case,
$\mu(\beta) = X \beta$, using the least squares estimator $\hat \beta =
(X^\prime X)^{-1} X^\prime Y$, we have $\mu(\hat \beta) = X \hat \beta
= H Y$, where $H = X (X^\prime X)^{-1} X^\prime$. Assuming the columns
of~$X$ are not colinear, the matrices $H$ and $M=I-H$ project onto
orthogonal subspaces of dimensions $p$ and $n-p$. The occurrence of $p$
at the left in (\ref{Inequality.three}) is usually viewed as connected
with the decomposition $Y^\prime Y = Y^\prime H Y + Y^\prime M Y$ and
the fact that the projection matrix $H$ has rank $p$. For a projection
matrix, however, rank and trace are identical, but it is the trace
which actually matters.

To appreciate this, note that if $\hat \mu_i$ is \textit{any} quantity
determined independently of $Y_i^*$, then
\begin{equation}\label{appreciate.one}
E ( Y_i^* - \hat \mu_i)^2  = E ( Y_i^* - \mu_i)^2 + E ( \hat \mu_i -
\mu_i)^2  .\hspace*{-30pt}
\end{equation}
On the other hand,
\begin{eqnarray}\label{appreciate.two}
E ( Y_i - \hat \mu_i)^2
&  =& E ( Y_i - \mu_i)^2  + E ( \hat \mu_i-
\mu_i)^2 \nonumber\hspace*{-30pt}
\\[-8pt]\\[-8pt]
& &{}- 2 \operatorname{Cov} (Y_i, \hat \mu_i)  .\nonumber\hspace*{-25pt}
\end{eqnarray}
Taken together, and remembering that $E ( Y_i^*\,{-}\,\mu_i)^2\,{=}\,\break E ( Y_i -
\mu_i)^2$, these give
\begin{equation}\label{appreciate.three}
 E ( Y_i^* - \hat \mu_i)^2  =  E ( Y_i - \hat \mu_i)^2 + 2 \operatorname{Cov}
(Y_i, \hat \mu_i)  ,\hspace*{-30pt}
\end{equation}
and then summing over $i$ shows that the difference between the right-
and the left-hand sides of (\ref{Inequality}) is
\[
2 \sum_{i=1}^n \operatorname{Cov} ( Y_i, \hat \mu_i )  .
\]
Equating this with $2 p \sigma^2$ leads to the definition\vspace*{1pt}
\begin{equation}\label{effective.df.cov}
\mbox{effective d.f.} \equiv \frac{1}{\sigma^2} \sum_{i=1}^n \operatorname{Cov} (Y_i, \hat \mu_i )  .
\vspace*{1pt}
\end{equation}
The relations (\ref{appreciate.one}), (\ref{appreciate.two}) and
(\ref{appreciate.three}) hold for \textit{any} estimator. But if $\hat
\mu = H Y$, that is, for a \textit{linear} estimator, the covariances
$\operatorname{Cov} ( Y_i, \hat \mu_i )$ are just the diagonal elements of $H$,
so that\vspace*{1pt}
\begin{equation}\label{effective.df.trace}
\mbox{effective d.f.} = \frac{1}{\sigma^2}  \operatorname{trace} (H)  .\vspace*{1pt}
\end{equation}
For \textit{nonlinear} models, the (approximate) effective number of
degrees of freedom may be defined either via (\ref{effective.df.cov}),
or via (\ref{effective.df.trace}) if we use the trace of its locally
linear approximation $\mu(\hat \beta) \simeq \mu(\beta_0) +
H(Y-\mu(\beta_0) )$, with both of these definitions being justifiable
asymptotically in view of (\ref{Equality.one}) and
(\ref{Equality.two}), under the smoothness condition referred to there.\vspace*{1pt}

\subsection*{Example: $I \times J$ ANOVA}\vspace*{1pt}

To help fix
ideas, it is instructive to consider the optimization problem for the
(complete) quadratically penalized $I \times J$ ANOVA\footnote{Unlike
the SVD case, discussed in (\ref{SVD.ALS.regular}) and in
footnote~\ref{footnote.lambda} of Section \ref{Section.SVD}, using different
values for $\lambda_1$ and $\lambda_2$ is essential here.}\vspace*{1pt}
\begin{eqnarray}\label{Penalized.ANOVA}
&&\sum_{i=1}^I \sum_{j=1}^J  ( r_{i,j} - \mu - \alpha_i - \beta_j
 )^2 \nonumber
 \\
 [-7pt]\\[-7pt]
 &&\quad{}+
\lambda_1  \Biggl( \sum_{i=1}^I \alpha_i^2  \Biggr) + \lambda_2  \Biggl( \sum_{j=1}^J
\beta_j^2  \Biggr) .\nonumber\vspace*{1pt}
\end{eqnarray}
We deliberately do not penalize for $\mu$ here because $\mu$ is
typically \textit{known} to differ substantially from zero. We will use
the identity\vspace*{1pt}
\begin{eqnarray}\label{ANOVA.special.identity}
&&\sum_{i=1}^I \sum_{j=1}^J  ( r_{i,j} - \mu - \alpha_i - \beta_j
 )^2  \nonumber\hspace*{-25pt}\\[1pt]
 &&\quad =
  \sum_{i=1}^I \sum_{j=1}^J  [  r_{i,j} - r_{\bolds\cdot,\bolds\cdot} -
 (  r_{i , \bolds\cdot} -  r_{\bolds\cdot, \bolds\cdot}   )  -  (
r_{\bolds\cdot , j} -  r_{\bolds\cdot, \bolds\cdot}   )   ]^2\nonumber\hspace*{-25pt}
\\[-7pt]\\[-7pt]
&&\qquad{}+   I  J   ( \mu - r_{\bolds\cdot, \bolds\cdot}   )^2
+
 \sum_{i=1}^I J  [ \alpha_i - ( r_{i , \bolds\cdot} -  r_{\bolds\cdot,
\bolds\cdot} )  ]^2 \nonumber\hspace*{-25pt}
\\[1pt]
&&\qquad{} +
  \sum_{j=1}^J I  [ \beta_j - ( r_{\bolds\cdot , j} -  r_{\bolds\cdot,
\bolds\cdot} )  ]^2 ,\nonumber\hspace*{-25pt}
\end{eqnarray}
where the ``dots'' represent averaging. It differs from the standard
ANOVA identity, but is derived similarly, although it requires
$\sum_{i=1}^I \alpha_i = 0$ and $\sum_{j=1}^J \beta_j = 0$. Using
(\ref{ANOVA.special.identity}), the optimization problem
(\ref{Penalized.ANOVA}) separates, leading to the solutions
\begin{eqnarray}\label{Shrunk.Eqn.A}
\qquad\hat \mu &=& r_{\bolds\cdot, \bolds\cdot}  ,\nonumber
\\[1pt]
\hat \alpha_i &=& \frac{J}{J + \lambda_1}  ( r_{i , \bolds\cdot} -
r_{\bolds\cdot, \bolds\cdot} ) \quad
\mbox{and}
\\[1pt]
   \hat \beta_j &=& \frac{I}{I + \lambda_2}  ( r_{\bolds\cdot , j}
-  r_{\bolds\cdot, \bolds\cdot} )  .\nonumber
\end{eqnarray}

Optimal choices for the regularization parameters~$\lambda_1$ and
$\lambda_2$ in (\ref{Penalized.ANOVA}) are usually estimated by
cross-validation, however, here we wish to understand these
analytically. We can do this by minimizing Akaike's predictive
information criterion (AIC),
\[
\mathrm{AIC} = - 2 \log ( {\mathcal L}_\lambda ) + 2  \operatorname{df}(\lambda)  ,
\]
where ${\mathcal L}_\lambda$ is the value (under $\lambda$-regularization)
of the likelihood for the $\{ r_{i,j} \}$ at the MLE, and $\operatorname{df}(\lambda)$ is the effective number of degrees of freedom; here
$\lambda \equiv (\lambda_1 , \lambda_2)$. As we are in a Gaussian case,
with an RMSE perspective, this is (except for additive constants) the
same as Mallows' ${\rm C}_p$ statistic,
\[
\mathrm{C}_p = \frac{ \{ \mbox{residual sum of squares} \}_\lambda
}{\sigma^2}   +    2  \operatorname{df}(\lambda)  .
\]
Minimizing this will (for linear models) be equivalent to minimizing
the expected squared prediction error, defined as the rightmost term in
(\ref{Inequality}), or (for nonlinear models) to minimizing it
asymptotically. For further discussion of these points, see Chapter~7
of Hastie et al. (\citeyear{Hastie}).

Now, the effective number of degrees of freedom associated with
(\ref{Penalized.ANOVA}) can be determined by viewing the minimizing
solution to (\ref{Penalized.ANOVA}) as a linear transformation, $\hat r
= H_\lambda r$, from the vector $r$ consisting of the observations
$r_{i,j}$, to the vector $\hat r$ of fitted values~$\hat r_{i,j}$. The
entries of the matrix $H_\lambda$ are determined from the relation
$\hat r_{i,j} = \hat \mu + \hat \alpha_i + \hat \beta_j$, where~$\hat\mu$,
$\hat \alpha_i$ and~$\hat \beta_j$ are given at
(\ref{Shrunk.Eqn.A}). Thus, the effective number of degrees of freedom,
when penalizing by $(\lambda_1, \lambda_2)$, is found to be
\begin{eqnarray}\label{Est.ANOVA.df}
\qquad\hspace*{5pt}\operatorname{df}   &=&    \operatorname{trace}  H_\lambda     \nonumber
\\[-8pt]\\[-8pt]
&=&    1 + (I-1)
\frac{J}{J+\lambda_1} + (J-1) \frac{I}{I+\lambda_2}  .\nonumber
\end{eqnarray}

Next, for a given $\lambda_1$ and $\lambda_2$, the residual sum of
squares is
\begin{eqnarray*}
&&\sum_{i=1}^I \sum_{j=1}^J  \biggl[
r_{i,j} - r_{\bolds\cdot,\bolds\cdot}
- \frac{J}{J+\lambda_1} ( r_{i,\bolds\cdot} -
r_{\bolds\cdot,\bolds\cdot} )\\
&&\hspace*{41pt}\hphantom{\sum_{i=1}^I \sum_{j=1}^J  \biggl[} -\frac{I}{I+\lambda_2} ( r_{\bolds\cdot,j} -
r_{\bolds\cdot,\bolds\cdot} )  \biggr]^2 ,
\end{eqnarray*}
and this may be expanded as
\begin{eqnarray*}
&&\sum_{i=1}^I \sum_{j=1}^J  [ ( r_{i,j} - r_{\bolds\cdot,\bolds\cdot} ) - (
r_{i,\bolds\cdot} - r_{\bolds\cdot,\bolds\cdot} ) - ( r_{\bolds\cdot,j} -
r_{\bolds\cdot,\bolds\cdot} )  ]^2
\\
&&\quad{}+  J  \biggl( 1 -\frac{J}{J +\lambda_1}  \biggr)^2 \sum_{i=1}^I ( r_{i,\bolds\cdot} -
r_{\bolds\cdot,\bolds\cdot} )^2
\\
&&\quad{} +  I  \biggl( 1 -\frac{I}{I +\lambda_2}  \biggr)^2
\sum_{j=1}^J ( r_{\bolds\cdot,j} - r_{\bolds\cdot,\bolds\cdot} )^2 ,
\end{eqnarray*}
where the first of the three terms here may subsequently be ignored.

Hence, the $\mathrm{C}_p$ criterion we seek to minimize can be taken as
\begin{eqnarray*}
&&\frac{1}{\sigma^2}  \sum_{i=1}^I \sum_{j=1}^J  \biggl[
r_{i,j} - r_{\bolds\cdot,\bolds\cdot} - \frac{J}{J+\lambda_1} ( r_{i,\bolds\cdot} -
r_{\bolds\cdot,\bolds\cdot} )\\
&&\hspace*{41pt}\hphantom{\frac{1}{\sigma^2}  \sum_{i=1}^I \sum_{j=1}^J  \biggl[}{}
-\frac{I}{I+\lambda_2} ( r_{\bolds\cdot,j} -
r_{\bolds\cdot,\bolds\cdot} )  \biggr]^2
\\
&&\quad{}+   2  \biggl\{
 1 +
(I-1) \frac{J}{J+\lambda_1} + (J-1) \frac{I}{I+\lambda_2}  \biggr\}
\end{eqnarray*}
or, equivalently,
\begin{eqnarray*}
&&\frac{1}{\sigma^2}   \Biggl\{  J  \biggl( 1 -\frac{J}{J +\lambda_1}
 \biggr)^2 \sum_{i=1}^I ( r_{i,\bolds\cdot} - r_{\bolds\cdot,\bolds\cdot} )^2
 \\
 &&\hphantom{\frac{1}{\sigma^2}   \Biggl\{}{} +  I
 \biggl( 1 -\frac{I}{I +\lambda_2}  \biggr)^2 \sum_{j=1}^J ( r_{\bolds\cdot,j} -
r_{\bolds\cdot,\bolds\cdot} )^2  \Biggr\}
\\
&&\quad{}+   2  \biggl\{ (I-1) \frac{J}{J+\lambda_1} + (J-1) \frac{I}{I+\lambda_2}  \biggr\}
.
\end{eqnarray*}
The minimizations with respect to $J / (J+\lambda_1)$ and $I/(I +
\lambda_2)$ thus separate, and setting derivatives to zero leads to the
approximate solutions
\begin{eqnarray}\label{Optimal.Lambdas}
\qquad\lambda_1 &=&    \biggl\{ \frac{\sigma^2} { \sum_{i=1}^I ( r_{i,\bolds\cdot} -
r_{\bolds\cdot,\bolds\cdot} )^2  /(I-1)}   \biggr\}\quad   \mbox{and} \nonumber
\\[-8pt]\\[-8pt]
  \lambda_2 &=& \biggl\{
\frac{\sigma^2} { \sum_{j=1}^J ( r_{\bolds\cdot,j} - r_{\bolds\cdot,\bolds\cdot}
)^2  /(J-1) }   \biggr\}  .\nonumber
\end{eqnarray}
On substituting these into (\ref{Est.ANOVA.df}), we also see that under
the theoretically optimal regularization the effective number of
degrees of freedom for the ANOVA becomes
\begin{eqnarray*}
&&(I + J - 1)
\\
&&\quad{}  -   \biggl\{ \frac{I-1}{J} \frac{\sigma^2}{  (
1/(I-1)  ) \sum_{i=1}^I ( r_{i,\bolds\cdot} - r_{\bolds\cdot,\bolds\cdot} )^2
}
\\
&&\hphantom{\quad{} -   \biggl\{}{}+  \frac{J-1}{I} \frac{\sigma^2}{  ( 1/(J-1)  ) \sum_{j=1}^J
( r_{\bolds\cdot,j} - r_{\bolds\cdot,\bolds\cdot} )^2 }  \biggr\}  ;
\end{eqnarray*}
the expression in braces gives the reduction in degrees of freedom
which results under the optimal penalization. Equations
(\ref{Optimal.Lambdas}) and (\ref{Shrunk.Eqn.A}) may be interpreted as
saying that optimal penalization (or shrinkage) should be done
differentially by parameter groupings, with each group of (centered)
parameters shrunk in accordance with that group's variability (the
variances of the row and column effects here) relative to the
variability of error, and each parameter in accordance with its support
base (i.e., with the information content of the data relevant to its
estimation---here $I$ and $J$).

\subsection*{Empirical Bayes Viewpoint}

The preceding
computations may be compared with an empirical Bayes approach. For this
we will assume that $r_{i,j} = \mu + \alpha_i +\beta_j + e_{i,j}$, with
the $e_{i,j}$ being independent $N(0, \sigma^2)$ variables. For
simplicity, we assume that $\mu$ and $\sigma^2$ are known. On the
parameters, $\alpha_i$ and $\beta_j$, respectively, we posit
independent $N(0, \sigma_1^2)$ and $N(0, \sigma_2^2)$ priors, with
$\sigma_1^2$ and $\sigma_2^2$ being hyperparameters. Multiplying up the
$I + J + IJ$ normal densities for the $\alpha_i$, $\beta_j$ and
$r_{i,j}$, and again using (\ref{ANOVA.special.identity}), we can
complete squares and integrate out the $\alpha_i$ and $\beta_j$. This
leads to a likelihood function for $\sigma_1^2$ and $\sigma_2^2$ which,
to within a factor not depending on $\sigma_1^2$ and $\sigma_2^2$, is
given by{\fontsize{10.2}{12.2}{\selectfont{
\begin{eqnarray*}
&&\hspace*{-6pt} \biggl(\frac{\sqrt{2\pi}\sigma}{\sqrt{J \sigma_1^2 + \sigma^2}} \biggr)^I
\exp  \Biggl[   \biggl( - \frac{1}{2 \sigma^2}  \biggr)  \biggl(  J - \frac{J^2}{J + (\sigma^2
/ \sigma_1^2)}  \biggr)
\\
&&\hspace*{160pt}{}\cdot \sum_{i=1}^I  ( r_{i,\bolds\cdot} - r_{\bolds\cdot,\bolds\cdot}
)^2   \Biggr]
\\
&&\hspace*{-6pt}\quad{}\cdot    \biggl(\frac{\sqrt{2\pi}\sigma}{\sqrt{I \sigma_2^2 + \sigma^2}} \biggr)^J
\exp  \Biggl[   \biggl( - \frac{1}{2 \sigma^2}  \biggr)  \biggl( I - \frac{I^2}{I + (\sigma^2 /
\sigma_2^2)}  \biggr)
\\
&&\hspace*{155pt}{}\cdot\sum_{j=1}^J  ( r_{\bolds\cdot,j} - r_{\bolds\cdot,\bolds\cdot}
)^2  \Biggr]  ,
\end{eqnarray*}}}}%
and maximizing this leads to the estimates
\begin{eqnarray*}
\hat \sigma_1^2   &=&  \frac{1}{I}  \sum_{i=1}^I  ( r_{i,\bolds\cdot} -
r_{\bolds\cdot,\bolds\cdot}  )^2 - \frac{\sigma^2}{J}   \quad  \mbox{and}
\\
  \hat \sigma_2^2   &=&   \frac{1}{J}  \sum_{j=1}^J  (
r_{\bolds\cdot,j} - r_{\bolds\cdot,\bolds\cdot}  )^2 - \frac{\sigma^2}{I} .
\end{eqnarray*}
The resulting empirical Bayes Gaussian prior can thus be seen as being
essentially equivalent to the quadratically penalized optimization
(\ref{Penalized.ANOVA}) under the optimal choice
(\ref{Optimal.Lambdas}) for the penalty parameters $\lambda_1,
\lambda_2$.

\subsection*{Generalizing}

We begin with a few remarks on
the penalized sparse ANOVA
\begin{eqnarray}\label{Penalized.ANOVA.sparse}
&&\sum \sum_{\mathcal C}  ( r_{i,j} - \mu - \alpha_i - \beta_j )^2\nonumber
\\[-8pt]\\[-8pt]
&&\quad{} +
\lambda_1  \Biggl( \sum_{i=1}^I \alpha_i^2   \Biggr) + \lambda_2  \Biggl( \sum_{j=1}^J
\beta_j^2  \Biggr)  .\nonumber
\end{eqnarray}
This optimization can be carried out by EM or by gradient descent; it
has no analytical solution, but analogy with the complete case suggests
that the shrinkage rules
\[
\hat \alpha_i^{\mathrm{shrink}} = \frac{J_i}{J_i + \lambda_1}  \hat
\alpha_i     
\quad\mbox{and}\quad   \hat \beta_j^{\mathrm{shrink}} = \frac{I_j}{I_j + \lambda_2}
\hat \beta_j  ,
\]
where $\hat \alpha_i$ and $\hat \beta_j$ are the unpenalized estimates,
will be approximately optimal provided we again take $\lambda_1$ and
$\lambda_2$ as ratios of row and column variation relative to  error as
at (\ref{Optimal.Lambdas}). Koren (\citeyear{KorenTwo}) proposed the less accurate but
simpler penalization
\[
\hat \beta_j = \frac {\sum_{i \in I(j)} (r_{i,j} - \hat \mu) } {I_j +
\lambda_2}
\]
first, and then
\[
   \hat \alpha_i = \frac {\sum_{j
\in J(i)} (r_{i,j} - \hat \mu - \hat \beta_j) } {J_i + \lambda_1}  ,
\]
where $\hat \mu$ is the overall mean; typical values he used\footnote{Koren's values were targeted to fit the probe set. If the probe and
training sets had identical statistical properties, these values would
likely have been smaller: recall that in Section \ref{Section.ANOVA} we
obtained variances of 0.23 and 0.28 for the user and movie means, and
RMSE values slightly below 1, suggesting the approximate values
$\lambda_1 \approx \lambda_2 \approx 4$.} were $\lambda_1 = 10$ and
$\lambda_2 = 25$.

For more complex models, such as sparse SVD, the lessons here suggest
that penalties on parameter groupings should correspond to priors which
model the distributions of the groups. For Gaussian priors (quadratic
regularization) we then need estimates for the group variances. For SVD
we thus want estimates of the variances of each of the user and movie
features. We experimented with fitting SVDs using minimal
regularization---with features in descending order of importance---first
removing low usage users to better assess the true user
variation. Because free constants can move between corresponding user
and movie features, we examined products of the variances of
corresponding features. These do tend toward zero (theoretically, this
sequence must be summable) but appear to do so in small batches,
settling down and staying near some small value, before settling still
further, again staying a while, and so on. Our explanation for this is
that there soon are no obvious features to be modeled, and that batches
of features then contribute small, approximately equal amounts of
explanatory power. Such considerations help suggest protocols for
increasing regularization as we proceed along features. It is an
important point\vadjust{\goodbreak} that, in principle, the number of features may be
allowed to become infinite, as long as their priors tend toward
degeneracy sufficiently quickly.

Bell, Koren and Volinsky (\citeyear{BellFour})         
proposed a particularly interesting empirical Bayes regularization for
the feature parameters in SVD. They modeled user parameters as $u_i
\sim N(\mu , \Sigma_1)$, movie parameters as $v_j \sim N(\nu ,
\Sigma_2)$, and individual SVD-based ratings as $r_{i,j} \sim N (
u_i^\prime v_j , \sigma^2)$, with the natural assumptions on
independence. They fitted such models using an EM and a Gibbs sampling
procedure, alternating between fitting the SVD parameters and fitting
the parameters of the prior. See also Lim and Teh (\citeyear{Lim}).

\section{Temporal Considerations}              
\label{Section.Temporal}

This section addresses the temporal discordances between the Netflix
training and qualifying data sets. See, for example, Figures \ref{f1} and \ref{f5} of
Section \ref{Section.Data} for evidence of such effects. Peoples'
tastes---collectively and individually---change with time, and the
movie ``landscape'' changes as well. The specific user who submits the
ratings for an account may change, and day-of-week as well as seasonal
effects occur as well. Furthermore, the introduction (and evolution)
of\break
a~recommender system itself affects ratings. Here we provide a very
brief overview of the main ideas which have been proposed for dealing
with such issues, although to limit our scope, time effects are not
emphasized in our subsequent discussions. Key references here are Koren
(\citeyear{KorenOne,KorenThree}).

We first note that temporal effects can be entered into models in a
``global'' way. Specifically, the standard baseline ANOVA can be modified
to read
\[
r_{i,j}  = \mu + \alpha_i(t) + \beta_j(t) + e_{i,j}  .
\]
Here all effects are shown as functions which depend on time, but the
time arguments $t$ can (variously) represent chronological time, or can
represent a user-specific or a movie-specific time $t_i$ or $t_j$, or
even a~jointly indexed time $t_{i,j}$.

Time effects can also be incorporated into both SVD and kNN type
models. An example in the SVD case is the highly accurate model
\begin{eqnarray*}
{\hat r}_{ij} (t) &=& \mu  + \alpha_i (t) + \beta_j (t)
\\
&&{}+ v_j^\prime
 \biggl( u_i (t) + |J(i)|^{ - 1/2} \sum_{j' \in J(i)} C_{j'}  \biggr),
\end{eqnarray*}
referred to as ``SVD$++$'' by Koren (\citeyear{KorenThree}), and fit using both
regularization and cross-validation. Here the baseline\vadjust{\goodbreak} values $\alpha_i
(t)$ and $\beta_j (t)$, as well the user effects $u_i (t)$, are both
allowed to vary over time but---on grounds that movies are more
constant than users---the movie effects $v_j$ are not. The last sum
models feedback from the implicit information. Detailed proposals for
temporal modeling of the user and movie biases, and for the user SVD
factors, $u_i(t)$, as well as for modeling temporal effects in nearest
neighbor models may be found in Koren (\citeyear{KorenOne,KorenThree}).

\section{In Search of Models}          
\label{Section.New}

Examining and contrasting such models as ANOVA, SVD, RBM and kNN is
useful in a search for new model classes. We first remark that the best
fitting models---such as SVD and RBM---have high-dimensional,
simultaneously fitted parameterizations. On the other hand, useful
models need not have, with ANOVA and kNN both suggestive of this. If
a~model has $p$ parameters, and if it is viewed as spanning a
$p$-dimensional submanifold of $R^N$, then we want $p$ to not be too
large, and yet for this submanifold to contain a vector close to the
expected $N$-dimensional vector of data to be fitted. For this to happen,
the model will have to reflect some substantive aspect of the structure
from whence the data arose. One striking feature of collaborative
filtering data is the apparent absence of any \textit{single} model that
can explain most of the explainable variation observed. The reason for
this may be that the available data are insufficient to reliably fit
such a model. Were sufficient data available, it is tempting to think
that some variation of SVD might be such a single model. In this
section we indicate some extensions to the models already discussed.
Most of these were arrived at independently, although many do contain
features resembling those in models proposed by others. It is to be
understood that regularization is intended to be used with most of the
procedures discussed.

\subsection*{Extending ANOVA}

Likert scales, such as the
Netflix stars system, are subjective, with each user choosing for
themselves what rating (or ratings distribution) corresponds to an
average movie, and just how much better (or worse) it needs to be to
change that rating into a~higher (or a lower) one. This not only speaks
to a~centering for each user, captured by $\alpha_i$ terms, but also to
a scaling specific to each user, and suggests a variation of the usual
ANOVA of the form
\begin{equation}\label{ANOVA.int.one}
r_{i,j} = \mu + \alpha_i + \gamma_i \beta_j + \mbox{error} .
\end{equation}
The scaling factors $\gamma_i$ here are meant to be shrunk toward 1 in
regularization. This model is of an interaction type, and may be
generalized to
\begin{equation}\label{ANOVA.int.two}\qquad
r_{i,j} = \mu + \alpha_i + \beta_j + \mbox{Interact} + \mbox{error}  ,
\end{equation}
where the Interact term in (\ref{ANOVA.int.two}) can vary
among\footnote{We do not mention $r_{i,j} = \mu + \alpha_i + \beta_j +
\gamma_i \beta_j$ which is equivalent to (\ref{ANOVA.int.one}), nor do
we include $\gamma_i \delta_j$ or, equivalently, $\gamma_i  \delta_j
\alpha_i \beta_j$, in (\ref{ANOVA.int.three}), as these are just
single-feature SVDs with baseline. However, we mention here the model
$r_{i,j} = \beta_j + \gamma_j \alpha_i$ which is a sister to
(\ref{ANOVA.int.one}), but has no convincing rationale behind it. Note
also that within the forms (\ref{ANOVA.int.three}), the $\alpha_i$
could be changed to $| \alpha_i |$ or $\alpha_i^2$, and similarly for
the $\beta_j$.}
\begin{eqnarray}\label{ANOVA.int.three}
&&\mbox{(a) } \gamma_i \alpha_i \beta_j , \quad  \mbox{(b) }  \gamma_j \alpha_i
\beta_j , \nonumber
\\[-8pt]\\[-8pt]
&&  \mbox{(c) }  \gamma_i  \delta_j \alpha_i    \quad\mbox{or}\quad   \mbox{(d) }
\gamma_i  \delta_j \beta_j .\nonumber
\end{eqnarray}
While these are all interaction models, note that (\ref{ANOVA.int.one})
is equivalent to a~one-feature SVD with only a~user baseline. Likewise,
note that~(\ref{ANOVA.int.three})(a)\break and~(\ref{ANOVA.int.three})(b) can be viewed as truncated
nonlinear SVDs. As an experiment, we fitted (\ref{ANOVA.int.one}) and
obtained an RMSE of 0.90256 on the training set as compared with 0.9161
from the 2-way ANOVA fit of Section~\ref{Section.ANOVA}. In terms of
MSE, this reduction is more than 6 times that expected under pure
randomness.

Finally, we remark that ANOVA ideas can also be adapted to model
probability distributions of ratings. A typical model of this type, in
obvious notation, is
\[
P [r_{i,j} = k]\propto\exp\bigl\{ \mu^{(k)}+\alpha_i^{(k)}+\beta_j^{(k)}  \bigr\}.
\]
If we wish, the dependence of $\alpha_i^{(k)}$ on $k$ here could be
suppressed. The numerical issues which arise here are similar to those
of the SVD-based multinomial model described below.

\subsection*{Extending SVD}

Likert scales are not
intrinsically linear; the distance between a 1 and 2 rating, for
instance, is not equivalent to the distance between a 4 and 5. This
suggests that the five possible rating values might first be
transformed into five other numbers, $g(r) = g_1 , g_2 , g_3, g_4$ and
$g_5$, say. Since SVD is scale but not location invariant, such
transformation offers 4 degrees of freedom. The $r_{i,j}$ can thus be
transformed into new data, $g_{i,j}$, say, and an SVD fitted to the
$g_{i,j}$ resulting in estimates $\hat g_{i,j} = u_i^\prime v_j$. These
fits may then be transformed back to the original scale by fitting a
transformation $\hat r_{i,j} = h(u_i^\prime v_j)$.

A further nonlinear extension to SVD is arrived at by arguing that
people and movies are not comparable entities,\vadjust{\goodbreak} so requiring their
descriptors to have equal lengths is artificial. Furthermore, users are
much more numerous than movies, so movie features are easier to
estimate, while user features create the more severe overfitting. If we
posit that each user is governed by $p$ features $u_i = (u_{i,1},
u_{i,2}, \ldots , u_{i,p})$, and each movie by $q$ features $v_j =
(v_{j,1}, v_{j,2}, \ldots , v_{j,q})$, with $p < q$, we may propose
models such as
\begin{eqnarray}\label{SVD.expansion}
r_{i,j} &=& \mu + \alpha_i + \beta_j + \sum_{k=1}^p \sum_{\ell=1}^q
a_{k,\ell} u_{i,k} v_{j,\ell} \nonumber
\\[-2pt]
&&{}+
\sum_{k=1}^p \sum_{k' = 1}^p\sum_{\ell=1}^q
b_{k,k',\ell} u_{i,k} u_{i,k'} v_{j,\ell}\nonumber
\\[-2pt]
&&{}+
\sum_{k=1}^p \sum_{\ell=1}^q \sum_{\ell' =1}^q
c_{k,\ell,\ell'} u_{i,k} v_{j,\ell} v_{j,\ell'}
\\[-2pt]
&&{}+
\sum_{k=1}^p \sum_{k' =1}^p  \sum_{\ell=1}^q \sum_{\ell' =1}^q
 d_{k, k', \ell, \ell'} u_{i,k} u_{i,k'} v_{j,\ell} v_{j,\ell'}\nonumber
\\[-2pt]
&&{}+ \mathrm{error}  .\nonumber
\end{eqnarray}
Such models can allow for additional flexibility, and modest gains from
the lower-dimensional parameterization combined with the reduced
regularization required.

SVD can also be adapted to model the multinomial distributions of the
$r_{i,j}$, instead of just their expected values. A typical model of
this type is
\begin{equation}\label{Probability.Model.two}
P  [  r_{i,j}  = k   ] = \frac { \exp  \{ u_i^\prime v_j^k
 \}   } { \sum_{\ell=1}^K \exp  \{ u_i^\prime v_j^\ell
 \}   }  ;
\end{equation}
here each movie $j$ is associated with five feature vectors $v_j^k$,
one for each rating value $k$. Note that, except for the absence of
ratings-dependent movie biases, (\ref{Probability.Model.two}) is
similar to the defining equation (\ref{Hinton.One}) of the RBM model.
Because movies are relatively few compared to users, the
parameterization of such models is not much greater than for a standard
SVD. Furthermore, the user terms $u_i$ in (\ref{Probability.Model.two})
can be modeled as sums of movie parameters, as indicated further below.
We remark that in one of our experiments, we tried to fit such models
solely using means and RMSE criteria, as in
\[
\operatorname{\sum\sum}\limits_{(i,j) \in \mathcal C}  \biggl(  r_{i,j}  - \frac{ \sum_{k=1}^K k \exp
 \{ u_i^\prime v_j^k   \}  } { \sum_{\ell=1}^K \exp  \{
u_i^\prime v_j^\ell   \} }  \biggr) ^2   +   \mathrm{penalty}  ,
\]
but encountered difficulties with convergence.

\subsection*{Deeper kNN}

The kNN models of Section
\ref{Section.kNN} can loosely be described as one layer deep; they
involve like-minded users\vadjust{\goodbreak} and/or similarly rated movies. However, this
does not exhaust the combinatorial possibilities. To focus on one
simple case, suppose user $i$ has seen movies, $j$ and $j'$, and that
we wish to predict his or her rating for movie $j''$. The remaining
users can then be partitioned into eighteen sets, according  to whether
they did or did not see each of $j$, $j'$ and~$j''$, and if they had
seen either of $j$ or $j'$, according to whether their ratings did or
did not agree with~$i$. Such partitioning can carry information
relevant to modeling $i$'s rating for $j''$, but we do not pursue these
issues here.

\subsection*{Lessons of the RBM}

We start by recalling the
defining equations (\ref{Hinton.One}) and (\ref{Hinton.Two}) for the
RBM model in the form
\begin{eqnarray}\label{Hinton.One.Ours}
&&P  (  r_{i,j} = k  |  h_i   )  \nonumber
\\[-8pt]\\[-8pt]
&&\quad=  \frac{\exp  ( b_j^k + \sum_{\ell=1}^F
h_{i,\ell} W_{j,\ell}^k  )} {\sum_{n = 1}^{K} \exp  ( b_j^n +
\sum_{\ell=1}^F h_{i,\ell} W_{j,\ell}^n  ) }\nonumber
\end{eqnarray}
and
\begin{equation}\label{Hinton.Two.Ours}
\quad P  (  h_{i,\ell} = 1  |  r_i   )  =  \sigma  \biggl( b_\ell + \sum_{j^\prime
\in J(i)} W_{j^\prime,\ell}^{r_{i,j^\prime}}  \biggr) .
\end{equation}
Examining these equations leads to valuable insights. First, the fact
that (in this version of the model) the hidden user features are
restricted to being binary seems inessential, except possibly in
contributing to regularization. In any case, binary features are
associated with probabilities, so users are, in effect, being described
by continuously-valued quantities. Second, it is not clear what
essential \textit{data-fitting} advantage is offered by viewing the user
features as being stochastic; indeed, the probabilities associated with
them may themselves be regarded as nonstochastic descriptors. (It may
be, however, that this randomness proxies an underlying empirical Bayes
mechanism.) On the other hand, having a probability model for the
$r_{i,j}$ seems natural---and perhaps even essential---for viewing
the data in a fuller context. Next, aside from its contribution to
parsimony, and to simplifying the fitting algorithms, the obligatory
symmetry of the $W_{j,\ell}^k$ weights\vspace*{1.5pt} in (\ref{Hinton.One.Ours})
and~(\ref{Hinton.Two.Ours}) seems restrictive. Finally, we remark that the
limitation on the bias terms $b_j^k$ in (\ref{Hinton.One.Ours}) to
depend on movie but not on user also seems restrictive. The RBM model
does offer certain advantages; in particular, it is trainable.

It pays to consider in further detail what it is that the RBM
equations, (\ref{Hinton.One.Ours}) and (\ref{Hinton.Two.Ours}),
actually do. The second of these equations, in effect, models each
user's features\vadjust{\goodbreak} as a function of the movies he or she has seen,
together with the ratings they had assigned to those movies. Doing so
limits the dimension of the parameterization for the user features, a
highly desirable goal. On the other hand, aside from the $b_j^k$ bias
terms, and aside from the stochastic nature of the user features, the
first equation models each of the multinomial probabilities for the
$r_{i,j}$ as a function of an SVD-like inner product of the user's
feature vector with a movie features vector (associated with the rating
value $k$) whose probability is being modeled.

Such considerations lead us to propose a model which we arrived at by
adapting the RBM equations~(\ref{Hinton.One.Ours}) and~(\ref{Hinton.Two.Ours})
for $P[r_{i,j} = k | h_i]$ and for\break
$P[h_{i,\ell} = 1 | r_i]$ into analogous equations for
expectations, namely,
\begin{equation}\label{E.r}
E  ( r_{i,j}  |  h_i  ) = g  \Biggl( b_j + \sum_{\ell=1}^F h_{i,\ell} \tilde
W_{j,\ell}  \Biggr)
\end{equation}
and
\begin{equation}\label{E.h}
E  ( h_{i,\ell} | r_i  ) = \tilde b_\ell + \sum_{j^\prime \in J(i)}
W_{j^\prime,\ell} ( r_{i,j^\prime})  .
\end{equation}
Here we have separated the different roles for the weights by using a
tilde in (\ref{E.r}); and because (\ref{E.r}) now models expectations
rather than probabilities, the dependence of the weights on $k$ there
has been removed. We next propose to use the right-hand side of
(\ref{E.h}) to estimate $h_{i,\ell}$, and substitute it into~(\ref{E.r}); the bias terms then all combine, and we are led to the
single equation model
\[
E  ( r_{i,j}  ) = g  \Biggl( b_j + \sum_{\ell=1}^F  \biggl\{ \sum_{j^\prime \in
J(i)} W_{j^\prime,\ell} ( r_{i,j^\prime})  \biggr\} \tilde W_{j,\ell}  \Biggr)
\]
or, generalizing this slightly,
\begin{eqnarray}\label{RBM.consequence}
\\[-6pt]
\hspace*{-1pt}
E  ( r_{i,j}  )
\,{=}\,g  \Biggl( b_j
 \,{+}\, \mbox{weight}
  \sum_{\ell=1}^F \biggl\{
\sum_{j^\prime \in J(i)}\! W_{j^\prime,\ell}(r_{i,j^\prime}) \biggr\} \tilde
W_{j,\ell}  \Biggr) .\nonumber
\end{eqnarray}
This model has SVD-like weights (features) $\tilde W_{j,\ell}$ for the
movies, and it models each user's weights (features) as a function of
the movies they have seen, using weights associated with the movies,
but depending also on the user's ratings for those movies. Although
derived independently, we note that Paterek (\citeyear{Paterek}) proposed a related
model, except that in Paterek's model, the movie functions which
determine the user weights [corresponding to our $W_{i,j}(k)$ here] do
not depend on~$k$, that is, on the user's ratings.\vadjust{\goodbreak} Our model is
therefore more general, but having more parameters requires different
penalization. See also the section on asymmetric factors in Bell, Koren
and Volinsky (\citeyear{BellOne}).

\subsection*{Modeling Users via Movie Parameters}

Parsimony of parameterization is a critical issue. Because users are 27
times more numerous than movies, assigning model parameters to users
exacts a far greater price, in degrees of freedom, than assigning
parameters to movies. This suggests that parameterization be arranged
in such a way that its dimension is a multiple of the number of movies
rather than of the number of users; we thus try to model user features
in terms of parameters associated with movies. In the context of the
Neflix contest, Paterek (\citeyear{Paterek}) was the first to have publicly suggested
this.

However, users cannot be regarded as being alike merely for having seen
the identical set of movies; their ratings for those movies must also
be taken into account. Such considerations lead to three immediate
possibilities:
\begin{longlist}[(A)]
\item[(A)] There is \textit{one} feature vector, $v_j$,
associated with each movie. The feature, $u_i$, for the $i$th
user is modeled as a sum (variously weighted) of functions of the $v_j$
for the movies
he or she has seen, and the ratings he or she assigned to them.
\item[(B)]
There are \textit{two} feature vectors, $v_j$ and $\tilde v_j$,
associated with each movie, and $u_i$ is based on the $\tilde v_j$ for
the movies $i$ has seen,
together with the ratings assigned to them.
\item[(C)] There are \textit{six} feature vectors, $v_j$ and $\tilde v_j^k
\equiv \tilde v_j (k)$, for $k = 1, 2, \ldots , K$ (with $K = 5$)
associated with each movie, and $u_i$ is based on the $\tilde v_j^k$
for the movies $i$ has seen,
and the ratings $k$ assigned to them.
\end{longlist}
(There are possibilities beyond just these three.)

These considerations lead to models such as
\[
\hat r_{i,j} = \mu + \alpha_i + \beta_j + u_i^\prime v_j  ,
\]
where the $v_j$ are free $p$-dimensional movie parameters, while the
$u_i$ are defined in terms of other ($p$-di-\break mensional) movie parameters.
For the approaches (A), (B) and (C) mentioned above, typical
possibilities include
\begin{eqnarray*}
u_i &=& \gamma \times \sum_{j^\prime \in J(i)}  ( r_{i,j^\prime} -
r_{i,\bolds\cdot}  ) v_{j^\prime}  ,
\\
   u_i &=& \gamma \times \sum_{j^\prime
\in J(i)} r_{i,j^\prime} \tilde v_{j^\prime}  \quad   \mbox{and}
\\
   u_i &=&
\gamma \times \sum_{j^\prime \in J(i)} \tilde v_{j^\prime} (r_{i,j})  ,
\end{eqnarray*}
respectively, where the $\gamma$'s are normalizing factors. In case
(C), for example, the overall model would become
\begin{eqnarray}\label{Overall.Model}
\quad\hat r_{i,j} &=& \mu + \alpha_i + \beta_j \nonumber
\\[-8pt]\\[-8pt]
&&{}+ \gamma \times \sum_{\ell =
1}^p \sum_{j^\prime \in J(i)} \tilde v_{j^\prime ,
\ell}(r_{i,j^\prime}) v_{j,\ell}  .\hspace*{-7pt}\nonumber
\end{eqnarray}
Note that (\ref{Overall.Model}) is essentially the same as the
RBM-inspired model (\ref{RBM.consequence}), but alternately arrived at.
Here the weights $\gamma$ might take the form $|J(i)|^{-\delta}$, with
typical possibilities for $\delta$ being $0$, $1/2$ or $1$, or as
determined by cross-validation.

\section{Further Ideas and Methods}            
\label{Section.Other}

Whether driven by fortune or by fame, the tenacity of the contestants
in the Netflix challenge has not often been surpassed. In this section
we collect together a few ideas which have not yet been discussed
elsewhere in this paper. A very few of these are our own (or at least
were obtained independently), but the boundaries between those and the
many other methods that have been proposed are necessarily blurred.

We start by noting that covariates can be included with many
procedures, as in
\begin{eqnarray*}
\hat r_{i,j} &=& \mu + \alpha_i(t) + \beta_j(t)
\\
&&{}+ \sum_{\ell=1}^p
u_{i,\ell} v_{j,\ell} + \sum_{m=1}^M c_m X_{i,j}^m + \cdots,
\end{eqnarray*}
where the $X_{i,j}^m$ for $m = 1, 2, \ldots , M$ are covariates. Such
models can generally be fit using gradient descent, and since
regularization is typically used as well, it would control
automatically for collinearities among covariates. Covariates introduce
very few additional parameters; hence, as a general rule (for this
data), the more the better. A covariate will typically differ across
both users and movies, unless it is viewed as being explanatory to the
``row'' or ``column'' effects. There are many possibilities for covariates.
(See, e.g., Toscher and Jahrer, \citeyear{Toscher}). A~selection of these
include the following:
\begin{enumerate}[9.]
\item[1.]  User and movie supports $J_i$, $I_j$ and various
        functions (singly and jointly) of these.
\item[2.]  Time between movie's release date and date rating was made.
\item[3.]  The number and/or proportion of movies
        user~$i$ has rated before and/or after rating movie $j$;
        the number and/or proportion of users who have rated movie $j$
        before and/or after user $i$ has rated it;
        and functions of these.\vadjust{\goodbreak}
\item[4.]  The relative density of movie ratings
        by the user around the time the rating was made;
        also, the movie's density of being rated around the time of rating.
\item[5.]  Seasonal and day-of-week-effects.
\item[6.]  Standard deviations and
variances of the user's ratings
        and of the user's residuals. Same for movies.
\item[7.]  Measures of ``surge of interest'' to detect ``runaway movies.''
\item[8.]  The first few factors or principal components
        of the covariance matrix for the movie ratings.
\item[9.]  Relationship between how frequently rated versus how highly
        rated the movie is,
        for example, $ ( \log I_j - {\rm Avg}  ) \times \beta_j$.
\end{enumerate}

We next remark that although they are the easiest to treat numerically,
neither the imposed RMSE criterion, nor the widely used quadratic
penalties, are sacrosanct for collaborative filtering. A mean absolute
error criterion, for instance, penalizes large prediction errors less
harshly or, equivalently, rewards correct estimates more generously,
and penalties based on $L^1$ regularization, as in the lasso
(Tibshirani, \citeyear{Tibshirani}), produce models with fewer nonzero parameters.
Although the lasso is geared more to model identification and parsimony
than to optimal prediction, the additional regularization it offers can
be useful in models with large parameterization. Other departures from
RMSE are also of interest. For example, if we focus on estimating \textit{probability distributions} for the ratings, a question of interest
is: With what probability can we predict a rating value \textit{exactly}? An objective function could be based on trying to predict
the highest proportion of ratings exactly correctly. A yet different
approach can be based on viewing the problem as one of ranking, as in
Cohen, Schapire and Singer (\citeyear{Cohen}). See also Popescul et al. (\citeyear{Popescul}).

A natural question is whether or not it helps to shrink estimated
ratings toward nearby integers.\break Takacs et al. (\citeyear{TakacsTwo}) considered this
question from an RMSE viewpoint and argued that the answer is no. One
can similarly ask whether it helps to shrink estimated ratings toward
corresponding modes of estimated probability distributions; we would
expect there too the answer to be negative.

Collaborative filtering contexts typically harbor substantial \textit{implicit} information. In many contexts, a user's search history or
even mouse-clicks can be useful. As mentioned several times previously,
for Netflix, \textit{which} movies a user rated carries information
additional to the actual ratings. Here 99\% of the data is ``missing,''
but not ``missing at random'' (MAR). Marlin et al. (\citeyear{MarlinOne}) discuss the
impact of the MAR assumption for such data. Paterek (\citeyear{Paterek}) introduced
modified SVD models (called NSVD) of the type
\[
\hat r_{i,j} = \mu + \alpha_i + \beta_j + v_j^\prime  \biggl( u_i +
|J_i|^{-1/2} \sum_{j^\prime \in J(i)} y_{j^\prime}  \biggr),
\]
where the $y_j$ are secondary movie features intended to model the
implicit choices a user has made. The Netflix \textit{qualifying} data
set, for example, contains important information by identifying many
cases of movies that users had rated, even though those rating values
were not revealed. Corresponding to such information is an $I \times J$
matrix which can be thought of as consisting of 0's and 1's, indicating
which user-movie pairs had been rated regardless of whether or not the
actual ratings are known. This matrix is full, not sparse, and contains
invaluable information. All leading contestants reported that including
such implicit information in their ensembles and procedures made a
vital difference. Hu, Koren and Volinsky (\citeyear{Hu}) treat the issue of
implicit information in greater detail. See also Oard et al. (\citeyear{Oard}).

Measures of similarity based on correlation-like quantities were
discussed in Section \ref{Section.kNN}. Alternate similarity measures
can be constructed by defining distances between the feature vectors of
SVD fits. Implicit versions of similarity may also be useful. For
example, the proportion of users who have seen mov\-ie~$j$ is $|I(j)| /
I$, and who have seen movie $j'$ is $|I(j')| / I$. Under independence,
the proportion who have seen both movies should be about $ |I(j)|
|I(j')| / I^2$, but is actually $|I(j, j')| / I$. Significant
differences between these two proportions is indicative of movies that
appeal to rather different audiences.

Among the most general models which have been suggested, Koren (\citeyear{KorenOne})
proposed combining the\break SVD and kNN methodologies while allowing for
implicit information within each component, leading to models such as
\begin{eqnarray*}
\hat r_{i,j}   &= &   \mu + \alpha_i + \beta_j + v_j^\prime  \biggl( u_i +
|J(i)|^{-1/2} \sum_{j \in J(i)} y_j  \biggr)
\\
&&{}+     | N^k(j;i)  |^{-1/2} \sum_{j' \in N^k(j;i)} (r_{i,j'} - b_{i.j'})
W_{j,j'}
\\
&&{}  +     | R^k(j;i)  |^{-1/2} \sum_{j' \in R^k(j;i)} C_{j,j'}
,
\end{eqnarray*}
where the $b_{i.j'}$ are a baseline fit. Here the sum involving the
$y_j$ makes the $v_j^\prime u_i$ SVD component ``implicit\vadjust{\goodbreak} information
aware.'' The sets $N^k(j;i)$ and $R^k(j;i)$ represent neighborhoods
based on the explicit and implicit information, respectively, while the
last sum is the implicit neighborhood based kNN term. This model is
among the best that have been devised for the Netflix problem.

\section{In Pursuit of Epsilon: Ensemble~Methods}               
\label{Section.Combining}

Meeting the 10\% RMSE reduction requirement of the Netflix contest
proved to be impossible using any single statistical procedure, or even
by combining only a small number of procedures. BellKor's 2007 Progress
Prize submission, for instance, involved linear combinations of 107
different prediction methods. These were based on variations on themes,
refitting with different tuning parameters, and different methods of
regularization (quadratic, lasso and flexible normal priors). BellKor
applied such variations to both movie and user oriented versions of
kNN, both multinomial and Gaussian versions of RBM, as well as to
various versions of SVD. Residuals from global and other fits were
used, as were covariates as well as time effects. The 2008 Progress
Prize submission involved more of the same, blending over 100 different
fitting methods, and, in particular, modeling time effects in deeper
detail; the individual models were all fit using gradient descent
algorithms. Finally, the Grand Prize winning submission was based on a
complex blending of no fewer than 800 models.

Several considerations underpin the logic of combining models. First,
different methods pick up subtly different aspects of the data so the
nature of errors made by different models differ; combining therefore
improves predictions. Second, prediction methods fare differently
across various strata of the data, and user behavior across such strata
differs as well. For instance, users who rated thousands of movies
differ from those who only rated only a few. If regularized (e.g.,
ridge) regression is used to combine estimators, the data can be
partitioned according (say) to support (i.e., based on the $J_i$
and/or~$I_j$), and separate regressions fit in each set, with the ridge
parameters selected using cross-validation.\break A~third consideration is
related to the absence of any unique way to approach estimation and
prediction in complex highly parameterized models. In the machine
learning literature, ways of combining predictions from many versions
of many methods are referred to as ensemble methods and are known to be
highly effective.\vadjust{\goodbreak} (See, e.g., Chapter 16 of Hastie, Tibshirani and  Friedman, \citeyear{Hastie}.)
In fact, the Netflix problem provides a striking and quintessential
demonstration of this phenomenon.

Various methods for blending (combining) models were used to
significantly improve prediction performance in the Netflix contest and
are described in Toscher and Jahrer (\citeyear{Toscher}), Toscher, Jahrer and Bell (\citeyear{Toscherthree}) and
Toscher, Jahrer and Logenstein (\citeyear{Toscherfour}). These include kernel ridge regression blending,
kNN blending, bagged gradient boosted decision trees and neural net
blending. Modeling the residuals from other models provided useful
inputs, for example, applying kNN on RBM residuals. It was found that
linear blending could be significantly outperformed and that neural net
blending combined with bagging was among the more accurate of the
proposed methods. Essentially, individual models were fit on the
training data while blending was done on the probe set, as it
represented the distribution of user/movie ratings to be optimized
over. In their winning submission, Toscher, Jahrer and Bell (\citeyear{Toscherthree}) noted that
optimizing the RMSE of individual predictors is not optimal when only
the RMSE of the ensemble counts; they implemented sequential
fitting-while-blending procedures as well as ensembles-of-blends.
Further details may be found in the cited papers. Some general
discussion of ensemble methods is given in Hastie et al. (\citeyear{Hastie}), Chapter~16.
See also DeCoste (\citeyear{DeCoste}), and Toscher, Jahrer and Legenstein (\citeyear{Toscherfour}).

It should be noted that although the Netflix contest required combining
very large numbers of prediction methods, good collaborative filtering
procedures do not. In fact, predictions of good quality can usually be
obtained by combining a small number of judiciously chosen methods.

\section{Numerical Issues}     
\label{Section.Numerical}

The scale of the Netflix data demands attention to numerical issues and
limits the range of algorithms that can be implemented; we make a few
remarks concerning our computations. We used a PC with 8 GB of RAM,
driven by a 3 GH, four-core, ``Intel Core~2 Extreme X9650'' processor.
Our computations were mainly carried out using compiled C$++$ code called
within a 64 bit version of MatLab running on a 64 bit Windows machine.

For speed of computation, storing all data in RAM was critical.
Briefly, we did this by vectorizing the data in two ways: In the first,
ratings were sorted by user and then by movie within user; and in the
second, conversely. A separate vector carried the ratings\vadjust{\goodbreak} dates. Two
other vectors carried identifiers for the users and for the movies;
those vectors were shortened considerably by only keeping track of the
indices at which a new user or a new movie began. Ratings were stored
as ``single'' (4 bytes per data point, so 400 MB for all ratings) and
user number as ``Int32'' (long integer, using 400 MB). As not all
variables are required by any particular algorithm, and dates often
were not needed in our work, we could often compress all required data
into less than 1 GB of RAM. Takacs et al. (\citeyear{TakacsTwo}) also discuss methods to
avoid swapping data across ROM.

Except for the RBM, we implemented many of the methods discussed in
this paper, as well as many others proposed in the literature. In
general, we found that gradient descent methods (stopping when RMSE on
the probe set is minimized) worked effectively.  For SVD, for example,
one full pass through the training data using our setup took
approximately 3 seconds; thus, fitting a regularized SVD of rank~40
(which required approximately 4000--6000 passes\break through) took
approximately 4--6 hours.

\section{Concluding Remarks}
\label{Section.Conclusion}

The Netflix challenge was unusual for the breadth of the statistical
problems it raised and illustrated, and for how closely those problems
lie at the frontiers of recent research. Few data sets are available,
of such dimensions, that allow both theoretical ideas and applied
methods to be developed and tested to quite this extent. This data set
is also noteworthy for its potential to draw together such diverse
research communities. It is to be hoped that similar contributions
could be made by other such contests in the future.

In this paper, we discussed many key ideas that have been proposed by a
large number of individuals and teams, and tried to contribute a few
ideas and insights of our own. To provoke by trivializing, let us
propose that there is one undercurrent which underlies most of what we
have discussed. Thus, ANOVA/baseline values can all be produced by the
features of an SVD. Likewise, fits from an SVD can be used to define
kNN neighborhoods. And finally, the internal structure of an RBM is, in
essence, analogous to a kind of SVD. Hence, if there a single
undercurrent, it surely is the SVD; that, plus covariates, plus a
variety of other effects. What complicates this picture are the
dimensions of the problem and of its parameterization, together with
the ensuing requirements for regularization, and the difficulties
(particularly the inaccuracies) of the estimation.

For the Netflix problem, an interesting question to speculate on is:
What is the absolutely best attainable RMSE? At one time, the 10\%
improvement barrier seemed insurmountable. But the algorithms of the
winning and runner up teams ultimately tied to produce a 10.06\%
improvement (Test RMSE 0.8567) over the contest's baseline. When the
prediction methods of these two top teams is combined using a 50/50
blend, the resulting improvement is 10.19\% (Test RMSE 0.8555); see
\texttt{\href{http://www.the-ensemble.com}{http://}
\href{http://www.the-ensemble.com}{www.the-ensemble.com}}.

The Netflix challenge also raises new questions. Some of these have
already been under active research in recent years, while others pose
new questions of problems that had been thought of as having been
understood.
For example, in the context of data sets of this size, how can one deal
most effectively with optimization under nonconvexity, as occurs, for
instance, in very sparse SVD? Can better algorithms be \mbox{devised} for
fitting RBM models, for having them converge to global optima, and for
deciding on early stopping for regularization purposes?
Furthermore, \mbox{currently} available theoretical results for determining
optimal cross-validation parameters are based on contexts in which the
distributions of the training data and of the cases for which
predictions are required are the same. Can these \textit{theoretical}
results be effectively extended to cover cases in which the training
and test sets are not identically distributed?
The Netflix problem also highlights the value of further work to gain
still deeper understanding of issues and methods surrounding
penalization, shrinkage and regularization, general questions about
bagging, boosting and ensemble methods, as well as of the trade-offs
between model complexity and prediction accuracy. Related to this are
questions about choosing effective priors in empirical Bayes contexts
(particularly if the number of parameters is potentially infinite), and
of the consequences of choosing them suboptimally.
What, for example, are the trade-offs between using a regularized model
having a very large number of parameters, as compared to using a model
having still more parameters but stronger regularization? For instance,
if two SVD models are fit using different numbers of features, but with
penalization arranged so that the effective number of degrees of
freedom of both models is the same, can one deal \textit{theoretically}
with questions concerning which model is better?
And finally, can general guidelines be developed, with respect to
producing effective ensembles of predictors, which apply to modeling of
large data sets requiring extensive parameterization? Such questions
are among the legacies of the challenge unleashed by the Netflix
contest.

\section*{Acknowledgments}

The authors thank Netflix Inc. for
their scientific contribution in making this exceptional data set
public, and for conducting a remarkable contest. This work was
supported by grants from the Natural Sciences and Engineering Research
Council of Canada. The work of Y.H. was additionally supported by a
grant from Google Inc., and by a Fields-MITACS Undergraduate Summer
Research Award. The authors thank the referees for their thoughtful
feedback on our manuscript.


\end{document}